\documentclass[twocolumn,english]{IEEEtran}
\usepackage[T1]{fontenc}
\usepackage[latin9]{inputenc}
\usepackage{units}
\usepackage{mathtools}
\usepackage{amsmath}
\usepackage{amssymb}
\usepackage{graphicx}
\usepackage{wasysym}

\makeatletter

\providecommand{\tabularnewline}{\\}

\usepackage{babel}
\ifCLASSOPTIONcompsoc
\else
\fi

\usepackage{cite}

\global\long\def\url#1{\expandafter\string\csname #1\endcsname}

\global\long\def\ie{\textit{i.e.,} }
\global\long\def\eg{\textit{e.g.,} }

\makeatother

\usepackage{babel}
\begin{document}
\title{Electromagnetic Models for Passive Detection and Localization of Multiple
Bodies}
\author{Vittorio Rampa, Gian Guido Gentili, Stefano Savazzi, and Michele D'Amico
	    \thanks{ V. Rampa and S. Savazzi are with the Istituto di
		Elettronica, di Ingegneria dell'Informazione e delle Telecomunicazioni (IEIIT), 
		Consiglio Nazionale delle Ricerche (CNR), Milano, Italy,
		e-mail: vittorio.rampa, stefano.savazzi@ieiit.cnr.it.
		G. G. Gentili and M. D'Amico are with the Dipartimento di Elettronica,
		Informazione e Bioingegneria (DEIB), Politecnico di Milano, Milano, Italy,
		e-mail: gianguido.gentili, michele.damico@polimi.it. This preprint
		has been submitted to IEEE for possible publication.}}
\maketitle
\begin{abstract}
The paper proposes a multi-body electromagnetic (EM) model for the
quantitative evaluation of the influence of multiple human bodies
in the surroundings of a radio link. Modeling of human-induced fading
is the key element for the development of real-time Device-Free (or
passive) Localization (DFL) and body occupancy tracking systems based
on the processing of the Received Signal Strength (RSS) data recorded
by radio-frequency devices. The proposed physical-statistical model,
is able to relate the RSS measurements to the position, size, orientation,
and random movements of people located in the link area. This novel
EM model is thus instrumental for crowd sensing, occupancy estimation
and people counting applications for indoor and outdoor scenarios.
The paper presents the complete framework for the generic N-body scenario
where the proposed EM model is based on the knife-edge approach that
is generalized here for multiple targets. The EM-equivalent size of
each target is then optimized to reproduce the body-induced alterations
of the free space radio propagation. The predicted results are then
compared against the full EM simulations obtained with a commercially
available simulator. Finally, experiments are carried out to confirm
the validity the proposed model using IEEE 802.15.4-compliant industrial
radio devices.
\end{abstract}


\thispagestyle{empty}

\pagestyle{empty}

\section{Introduction}

\label{sec:Introduction}

\IEEEPARstart{H}{uman} Presence-aware Systems (HPSs) are rapidly
growing as new services become available in various areas of modern
life \cite{Petrov-2019} such as assisted living, ambient intelligence,
smart spaces, home automation, human-robot collaboration, safety and
security, just to cite a few. Among these applications, non-cooperative,
also known as passive, HPS are the most attractive since they do not
require the monitored users to carry or wear any electronic device
or specific sensors. Usually, these systems are vision-based \cite{dalal-2005,benezeth2011towards,choi-2013};
however, the ubiquitous presence of wireless networks paves the way
towards the exploitation of wireless radio-frequency (RF) networks,
not only as communication devices, but also as body proximity/location
virtual sensors. Last but non least, radio-based HPS are privacy-neutral
since they do not reveal any privacy information about the monitored
people.

HPS systems exploit the fact that people, or obstacles, in the surroundings
of an area covered by a wireless radio network induce signal alterations
that can be detected and exploited for body occupancy inference applications.
For instance, Device-Free Localization (DFL) systems \cite{Youssef:2007:CDP:1287853.1287880,savazzi2016magazine}
exploit a network of RF nodes to detect the presence, to locate and
track the position of moving objects or people in a confined area
covered by the wireless network itself. However, a radio-based HPS
is not only able to localize and track \cite{wilson10,nicoli2016eusipco,wang-2017}
people, or objects, but it has been also proven to efficiently perform
other tasks such as to count the number of people \cite{depatla2015jsac},
to identify and recognize patterns related to their activities \cite{savazzi2016icassp,wang2015icmcn}
and intentions \cite{savazzi2016icassp}, to detect dangerous worker
conditions and safety status \cite{kianoush2016iot,Talebpour-2019},
and act as a proximity monitor \cite{montanari2017proximity}. This
is made possible as the presence of targets (\ie objects or people)
affects the propagation of the radio waves in the covered area \cite{Woyach-et-al,patwari10},
for example by inducing predictable alterations of the Received Signal
Strength (RSS) field that depend on the targets position, in both
static \cite{Seifeldin-et-al} and dynamic \cite{Saeed-et-al} environments.

\subsection{Related works}

\label{subsec:Related-works}

The effect of the presence of people on the received RF signals is
a well know topic \cite{obayashi1998body,villanese2000pedestrian}
and finds its roots in the research activities about the electromagnetic
(EM) propagation phenomena caused by natural or artificial obstacles
during the first experimental trials at the dawn of the radio era
\cite{Brittain1994}. These studies have received a great impulse
after the middle of the last century mostly for outdoor coverage applications
\cite{furutsu-1963,vogler-1982,comparative,tzaras-2000}. However,
despite some recent attempts to model the body-induced fading effects
on short-range radio propagation \cite{smith2013propagation}, these
research activities are mostly related to inter- \cite{Koutitas,de2015analysis}
and intra-body \cite{Namjun_Cho-et-al,andreu2016spatial} short-range
radio communications. The aim of these research activities is to quantify
the radio propagation losses in narrow \cite{humanbody2} or wide-band
\cite{Fort-et-al} indoor scenarios with the main purpose of mitigating
these effects. Only a few research works \cite{Koutitas,liu2009fading,conducting_cylinder}
focus their attention on the geometrical relations between the transmitter
(TX) and receiver (RX) location, the position and composition of the
body, and its size.

A general EM model for the prediction of the mathematical relations
between location, size and composition of a \emph{single-target} and
the corresponding EM field perturbation, is still disputable as shown
in \cite{hamilton-et-al,yiugitler2016experimental} or too complex
to be of practical use as based on ray tracing techniques \cite{de2015analysis,eleryan-et-al}
or Uniform Theory of Diffraction (UTD) \cite{Koutitas,conducting_cylinder}.
Other EM methods \cite{kibret2015characterizing,ziri2005prediction,yokota2012propagation}
and physical-statistical models \cite{eleryan-et-al,mohamed2017physical,rampa2017em,Hillyard-2020}
are simpler than the previous ones but still limited to a single target.
To the authors knowledge, and according to the current literature,
an analytical, or semi-analytical, approach towards a \emph{true}
\emph{multi-body} model has never been tackled before. Usually, multi-target
(\ie multi-body) problems have been solved by assuming the linear
superposition of the single-body extra attenuations \cite{patwari10,wilson10}.
However, the mutual effects induced by multiple bodies moving concurrently
in the same space must be accounted for.

In \cite{nicoli2016eusipco} a DFL system has been proposed to track
two targets moving concurrently by using an EM model that is fully
described in \cite{Rampa-2019}. On the contrary, in this paper, the
model is generalized to an \emph{arbitrary number of targets}. A practically-usable
physical-statistical model is thus designed for the prediction and
the evaluation of the body-induced propagation losses, namely the
RSS field, found in true $N$-targets scenarios with $N\geq1$. This
$N$-body model is able to describe both dominant static component
and stochastic fluctuations of the power loss as a function of the
locations of the $N$ targets, their size, orientation and random
movements with respect to the link path.

\subsection{Original contributions}

\label{subsec:Original-contributions}

The paper proposes an EM framework where the field perturbations induced
by an arbitrary number of human bodies are modelled as a superposition
of \emph{diffraction} and\emph{ multipath} terms. The diffraction
component is defined according to the scalar diffraction theory and
is characterized by the geometrical description (\ie location, size,
orientation) and the movement characteristics (\ie rotations and
random movements around the nominal position) of $N$ targets according
to the knife-edge hypothesis \cite{lee1978path,vogler-1982,Edge_diffraction,deygout1991correction}.
The multipath fading term is assumed to impair the radio link due
to the presence of the bodies placed inside the sensitivity area \cite{rampa2015letter}
around the LOS (Line Of Sight) path that connects the transmitter
and the receiver. However, unlike \cite{rampa2015letter}, where RSS
perturbations are predicted for a \emph{single} \emph{small} target
\cite{comparative,rampa2015letter,knife_edge} moving only in the
central part of the LOS path according to the paraxial approximation
\cite{rampa2015letter}, this novel model provides a representation
of the power losses induced by \emph{multiple} bodies having \emph{any}
size, and placed \emph{anywhere} in the area surrounding the radio
link. The model presented here extents the dual-body case exploited
in \cite{nicoli2016eusipco} and then presented in \cite{Rampa-2019},
by considering a generic EM scenario with an arbitrary number of human
bodies in the surroundings of a radio link. In the former reference
\cite{nicoli2016eusipco}, the dual-body model is neither derived
nor justified but it is just introduced to perform DFL tasks and compare
the results against other methods. In the latter reference \cite{Rampa-2019},
the dual-body model is derived from prime principles and then described
and discussed in details. The experimental results presented here
confirm that the proposed model can effectively describe the mathematical
relations between the target positions and the measured RSS values.
Comparisons with the results obtained with the EM simulator Feko also
support the validity of the proposed model.

The novel contributions of this paper are: \emph{i}) the definition
of a general EM framework for the multi- body scenario; \emph{ii})
the derivation, from prime principles, of the full equations for the
prediction of the global extra attenuation due to \emph{$N$} bodies,
or objects, in the LOS area; \emph{iii}) the derivation of the analytical
formulas in the case of paraxial hypothesis for the general $N$ bodies
scenario; \emph{iv}) the evaluation of the extra attenuation predictions
for the dual-body scenario (\ie \emph{$N=2$}) and their comparison
against the results obtained using full EM simulations; and \emph{v})
tuning of the dual-body model parameters based on-field RSS measurement
trials and comparisons of the model predictions against the aforementioned
RSS measurements.

The paper is organized as follows. The diffraction model that accounts
for the deterministic term of the multi-body induced extra attenuation
is shown in Sect. \ref{sec:Diffraction-model} for any number $N$
of the targets. The complete physical-statistical model for the prediction
of the RSS field is illustrated in Sect. \ref{sec:Physical-statistical-modeling}.
In particular, the dual-body model is highlighted as a practical case
study. Sect. \ref{subsec:Model-calibration} deals with the evaluation
of the proposed multi-target model featuring a comparative analysis
against experimental measurements and simulation results. The concluding
remarks are drawn in Sect. \ref{sec:Conclusions}.

\begin{figure}[tp]
\begin{centering}
\includegraphics[clip,scale=0.35]{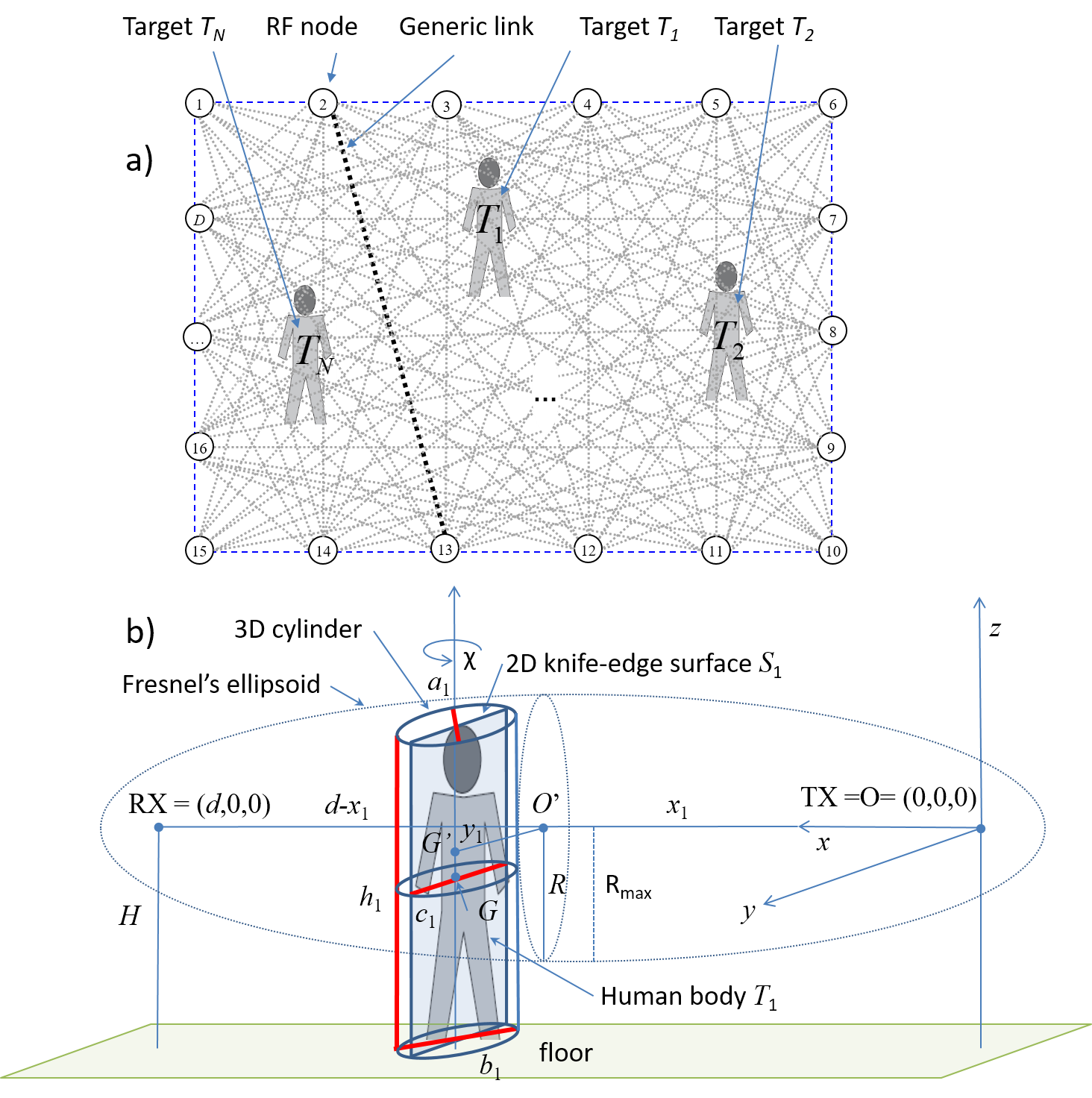}
\par\end{centering}
\caption{\label{fig:layout}a) Generic layout of a HPS-based wireless network
composed by $D$ nodes and $L$ links where $T_{n}$ is the $n$-th
target located inside the monitored area; b) Simplified representation
of the single-link single-body scenario where the human body $T_{1}$
is sketched as a 3D cylinder and then simplified as a 2D knife-edge
surface $S_{1}$.}

\vspace{-0.4cm}
\end{figure}

\section{Diffraction framework for the multi-body scenario}

\label{sec:Diffraction-model}

As sketched in Fig. \ref{fig:layout}.a, a generic HPS consists of
a mesh of partially, or completely connected, wireless network composed
of $D$ RF nodes \cite{Youssef:2007:CDP:1287853.1287880,savazzi2016magazine}
and $L\leq D\left(D-1\right)/2$ bidirectional links. The HPS-enabled
network is composed by nodes that are able to perform power measurements
on the RF signal and, after some processing steps, to extract body
occupancy-related information. We assume that all the RF nodes can
measure the RSS field values at discrete time instants. No specific
additional RF hardware \cite{savazzi2016magazine} is required since
RSS values are computed in the normal operations of the networked
RF nodes for channel estimation/equalization and frequency/frame synchronization
tasks.

Without any loss of generality, as described in Fig. \ref{fig:layout}.b,
in what follows, we will focus on the single-link scenario (\ie $L=1$
and $D=2$), by introducing the single-target ($T_{1}$ being $N=1$)
first and then the multi-target ($T_{1},...,T_{N}$ with $N>1$) cases.
However, the multi-body model presented here can be exploited in a
general multi-link scenario with $D$ nodes, $L$ links and $N$ targets
by using electric field superposition. In addition, it can be extended
to make use of other physical layer channel information measures such
as the Channel State Information (CSI) and the Channel Quality Information
(CQI) \cite{savazzi2016magazine} as well. However, this discussion
is outside the scope of this work.

It is worth noticing that all the proposed models apply to a generic
link of the radio network: therefore, they could be easily tailored
to predict RSS over arbitrarily complex network structures for more
robust body positioning, as proposed in device-free localization methods
\cite{patwari10,wilson10,savazzi2016magazine}. In addition, modeling
of RSS is instrumental for link selection operations, namely to identify
an optimized subset of links that are most influenced by the target
presence \cite{nicoli2016eusipco}.

\begin{figure}[tp]
\begin{centering}
\includegraphics[clip,scale=0.24]{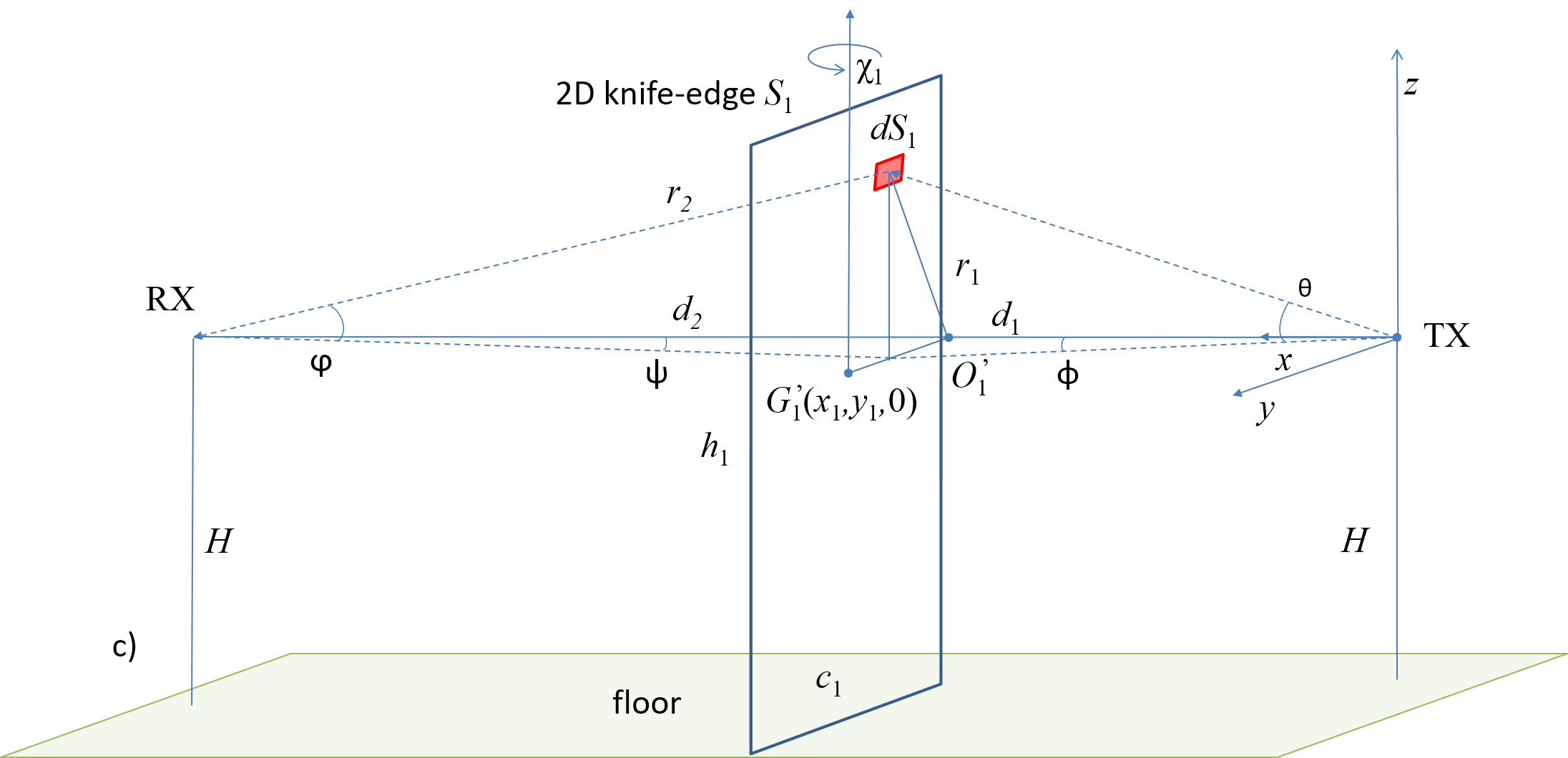}
\par\end{centering}
\caption{\label{fig:single}Geometrical representation of the same scenario
shown in \ref{fig:layout}.b, where an horizontal link of length $d$
is positioned at distance $H$ from the floor and a 2D knife-edge
surface $S_{1}$, with variable traversal size $c_{1}$ and height
$h_{1}$, represents the body $T_{1}$ that is placed on the floor.}

\vspace{-0.4cm}
\end{figure}

\subsection{Single body model (SBM)}

\label{subsec:Single-body-model}

The single body model \cite{rampa2017em} is briefly recalled in this
section since it is the starting point for the multi-target model
that will be described in Sect. \ref{subsec:Multi-body-model}. As
outlined in Fig. \ref{fig:single} , the human body (\ie the only
target $T_{1}$ located near the single-link area) is represented
by a perfectly EM absorbing 3D homogeneous cylinder with an elliptical
base of minor and major axes $a_{1}$ and $b_{1}$, respectively,
that simulate the human head, torso, legs and arms (placed near the
torso). Most references assume a 3D cylinder with a circular base
\cite{conducting_cylinder,ghaddar2004modeling} or a prism \cite{de2015analysis}
while only a few \cite{liu2009fading,reusens2007path} assume also
that the arms can freely move with respect to the torso. Considering
a dynamic scenario where the 3D cylinder, modeling the body, can freely
move horizontally and rotate around its generic \emph{nominal} position
$(x,y,z)$ showing different views, the target is reduced \cite{rampa2017em}
to a 2D rectangular blade (\ie a knife-edge surface) \cite{Edge_diffraction},
orthogonal to the LOS path at distance $d_{1}$ and $d_{2}$ from
the TX and RX, respectively. The knife-edge surface is vertically
placed close to the link area and can freely rotate and move showing
different body views during its movements. The presence of the floor
does not imply any influence on the EM propagation and it is used
here only for geometrical reasons \ie to define the height of the
link and the placement constraints of the knife-edge surface representing
the body. Notice that the knife-edge approximation ignores important
EM parameters such as polarization, permittivity, conductivity, shape,
radius of curvature, and surface roughness \cite{Davis-et-al}.

According to the Fig. \ref{fig:layout}.b, the radio link is horizontally
placed at distance $H$ from the floor and the 3D target $T_{1}$,
that is placed on the floor, is free to move and rotate around the
vertical axis in the surroundings of the LOS path. The corresponding
first Fresnel's zone ellipsoid \cite{Edge_diffraction}, with radius
$R=\sqrt{\lambda d_{1}d_{2}/d}$, does not have any contact with all
other parts of the scenario (\eg walls, ceiling, furniture or other
obstacles) except for the aforementioned target. Being $R\leq R_{\max}=\sqrt{\lambda d}/2$,
where $R_{\max}$ is the maximum value of the radius $R$, $\lambda$
is the wavelength and $d$ is the RX-TX distance (\ie the link path
length), this constraint becomes $2H>\sqrt{\lambda d}$. Notice that,
as stated by standard short-range indoor propagation models \cite{ITU-indoor},
ground attenuation effects may be safely ignored for a radio link
inside a single indoor large room/space \emph{e.g.} a hall.

The equivalent 2D knife-edge surface $S_{1}$ has height $h_{1}$,
width $c_{1}$, and it is placed orthogonal to the LOS path at location
$\mathbf{X}_{1}=[x_{1},y_{1}]^{T}$. $\mathbf{X}_{1}$ coincides with
the first two coordinates of the $S_{1}$ barycenter $G_{1}=\left(x_{1},y_{1},z_{1}\right)$
of the knife-edge $S_{1}$ since $z_{1}$ assumes the constant value
$z_{1}=h_{1}/2-H$. The point $G_{1}^{'}$ is the intersection of
the vertical axis passing through $G_{1}$ and the horizontal plane
$z=0$. In the followings, the position of the target $T_{1}$ (\ie
the position of $G_{1}^{'}$) is thus identified by the off-axis displacement
$y_{1}$ and the distance $x_{1}=d_{1}$ of $S_{1}$ from the TX.
However, a true person can also turn and make involuntary/voluntary
movements while standing on the floor. Therefore, the 3D target $T_{1}$,
represented by the 2D knife-edge surface $S_{1}$, can assume any
orientation $\chi_{1}\in[-\pi,\pi]$ with respect to the LOS path.
It can make also some small movements $\Delta\mathbf{X}_{1}=[\Delta x_{1},\Delta y_{1}]^{T}$
around the nominal location $\mathbf{X}_{1}$ thus showing a changing
traversal size $c_{1}=c_{1}(\chi_{1})\in[a_{1},b_{1}]$ and location
$\mathbf{X}_{1}+\Delta\mathbf{X}_{1}$.

According to the scalar theory of diffraction, the electric field
at the RX, that is generated by the isotropic source in TX is modified
by the presence of the 2D knife-edge surface $S_{1}$ located in the
link area \cite{rampa2017em}. It can be predicted \cite{comparative}
as being generated by a virtual array of Huygens' sources located
on $S_{1}$ but not belonging to the obstacle itself. In far field
conditions, the electric field $dE$ at the RX due to the diffraction
effects caused by the elementary Huygens' source of area $dS_{1}$
with generic coordinates $\left(x,y,z\right)$ is given by

\begin{equation}
dE=j\,\frac{E_{0}\,d}{\lambda\,r_{1}\,r_{2}}\,\exp\left\{ -j\frac{2\pi}{\lambda}\left(r_{1}+r_{2}-d\right)\right\} dS_{1}\,,\label{eq:dE_tot}
\end{equation}
where $r_{1}$ and $r_{2}$ are the distances of the generic elementary
area $dS_{1}$ for the TX and RX, respectively. $E_{0}$ is the free-space
electric field that is described by the following equation

\begin{equation}
E_{0}=-j\,\frac{\eta\,I\text{\ensuremath{\ell}}}{2\,\lambda\,d}\,\exp\left(-j\frac{2\,\pi\,d}{\lambda}\right)\,,\label{eq:E0}
\end{equation}
with $\eta$ being the free-space impedance and $I\text{\ensuremath{\ell}}$
the momentum of the source. The electric field at the RX is given
by \cite{rampa2017em}

\begin{equation}
\begin{array}{ll}
E= & -j\,\frac{\eta\,I\text{\ensuremath{\ell}}}{2\,\lambda\,d}\,\exp\left(-j\frac{2\,\pi\,d}{\lambda}\right)\cdot\\
 & \cdot\left\{ 1-j\,\frac{d}{\lambda}\int\limits _{S_{1}}\frac{1}{r_{1}\,r_{2}}\,\exp\left\{ -j\frac{2\pi}{\lambda}\left(r_{1}+r_{2}-d\right)\right\} dS_{1}\right\} \,,
\end{array}\label{eq:E_single}
\end{equation}
where the first term refers to the electric field (\ref{eq:E0}) due
the free-space propagation in the empty scenario and the second term
includes the diffraction effects due to the body presence according
to Eq. (\ref{eq:dE_tot}). The integral of the second term, is computed
over the rectangular domain defined by the $S_{1}$ region defined
as $S_{1}=\left\{ \left(x,y,z\right)\in\mathbb{R^{\textrm{3}}}:\:x=x_{1}=d_{1}\right.$,
$y_{1}-c_{1}/2\leq y\leq y_{1}+c/2$, $\left.-H\leq z\leq h_{1}-H\right\} $.
Focussing on the \emph{extra attenuation} induced by the body \emph{w.r.t.}
the free-space, Eq. (\ref{eq:E_single}) can be written as

\begin{equation}
\frac{E}{E_{0}}=1-j\,\frac{d}{\lambda}\int\limits _{S_{1}}\frac{1}{r_{1}\,r_{2}}\,\exp\left\{ -j\frac{2\pi}{\lambda}\left(r_{1}+r_{2}-d\right)\right\} dS_{1}\,.\label{eq:E_E0_tot-single}
\end{equation}
According to (\ref{eq:E_E0_tot-single}), the presence of the body
induces an extra attenuation $A_{\textrm{dB}}=-10\,\log_{10}\left|E/E_{0}\right|^{2}$
\emph{w.r.t.} the free-space propagation. Being a forward only method,
the diffraction model holds only for the generic target $T_{1}$ placed
in the area $\mathcal{Y}$ near the radio link where it is $\mathcal{Y}=\left\{ \left(x,y\right)\subset\mathbb{R^{\textrm{2}}}:0<x<d;-\infty<y<+\infty\right\} $.
Of course, as shown in \cite{rampa2017em}, the effect of the target
presence vanishes for large but finite values of $\left|y\right|$.

\subsection{Paraxial single body model (PSBM)}

\label{subsec:Paraxial-single-body}

By exploiting the paraxial approximation and the following variable
substitution $u=y\tfrac{\sqrt{2}}{R_{1}}$ and $v=z\tfrac{\sqrt{2}}{R_{1}}$,
Eq. (\ref{eq:E_E0_tot-single}) reduces to the PSBM (Paraxial Single
Body Model) equation defined as
\begin{equation}
\begin{aligned}\frac{E}{E_{0}}={} & 1-\frac{1}{2}\:j\varint_{\left(\sqrt{2}y_{1}-c_{1}/\sqrt{2}\right)/R_{1}}^{\left(\sqrt{2}y_{1}+c_{1}/\sqrt{2}\right)/R_{1}}\exp\left(-j\frac{\pi}{2}u^{2}\right)du\cdot\\
{} & \cdot\int_{-\unit{\sqrt{2}\mathit{H}/\mathit{R_{1}}}}^{+\sqrt{2}\left(h_{1}-H\right)/R_{1}}\exp\left(-j\frac{\pi}{2}v^{2}\right)dv\,,
\end{aligned}
\label{eq:E_E0_approx-single}
\end{equation}
where the Fresnel's radius $R_{1}=R\left(x_{1}\right)$ is given by
$R_{1}=\sqrt{\lambda\,x_{1}\,\left(d-x_{1}\right)/d}$. Here, we have
specified that the generic Fresnel's radius $R$ of Fig. \ref{fig:layout}.b
is equal to $R\left(x_{1}\right)$ to explicitly highlight that it
is related to the position $x_{1}$ of target $T_{1}$. Eq. (\ref{eq:E_E0_approx-single})
can be easily computed by using the Fresnel's Sine and Cosine Integrals
\cite{rampa2015letter}. The paraxial approximation implies the following
assumptions: $\max\{|y_{1}|,|z_{1}|,h_{1},c_{1},\lambda\}\ll\min\{x_{1},d-x_{1}\}$,
$\cos\varphi\simeq1$, $\cos\theta\simeq1$, $\cos\psi\simeq1$, and
$\cos\phi\simeq1$, where the aforementioned angles are shown in Fig.
\ref{fig:single}. The model gives valid predictions in the same area
$\mathcal{Y}$ defined for the SBM model. The paraxial approximation
is mainly used for outdoor scenarios, namely for terrestrial radio
propagation applications \cite{furutsu-1963,vogler-1982} where it
holds gracefully. However, it has been also employed in HPS applications
\cite{dfl}, even if this paraxial hypothesis limits the validity
of Eq. (\ref{eq:E_E0_approx-single}) for small bodies near the central
area of the radio link. For more details about single target modeling
derived from (\ref{eq:E_E0_tot-single}), \emph{e.g.} for models that
include vertical or horizontally polarization, the interested reader
can refer to \cite{rampa2017em}.

\subsection{Multi-body model (MBM)}

\label{subsec:Multi-body-model}

\begin{figure}[tp]
\begin{centering}
\includegraphics[clip,scale=0.24]{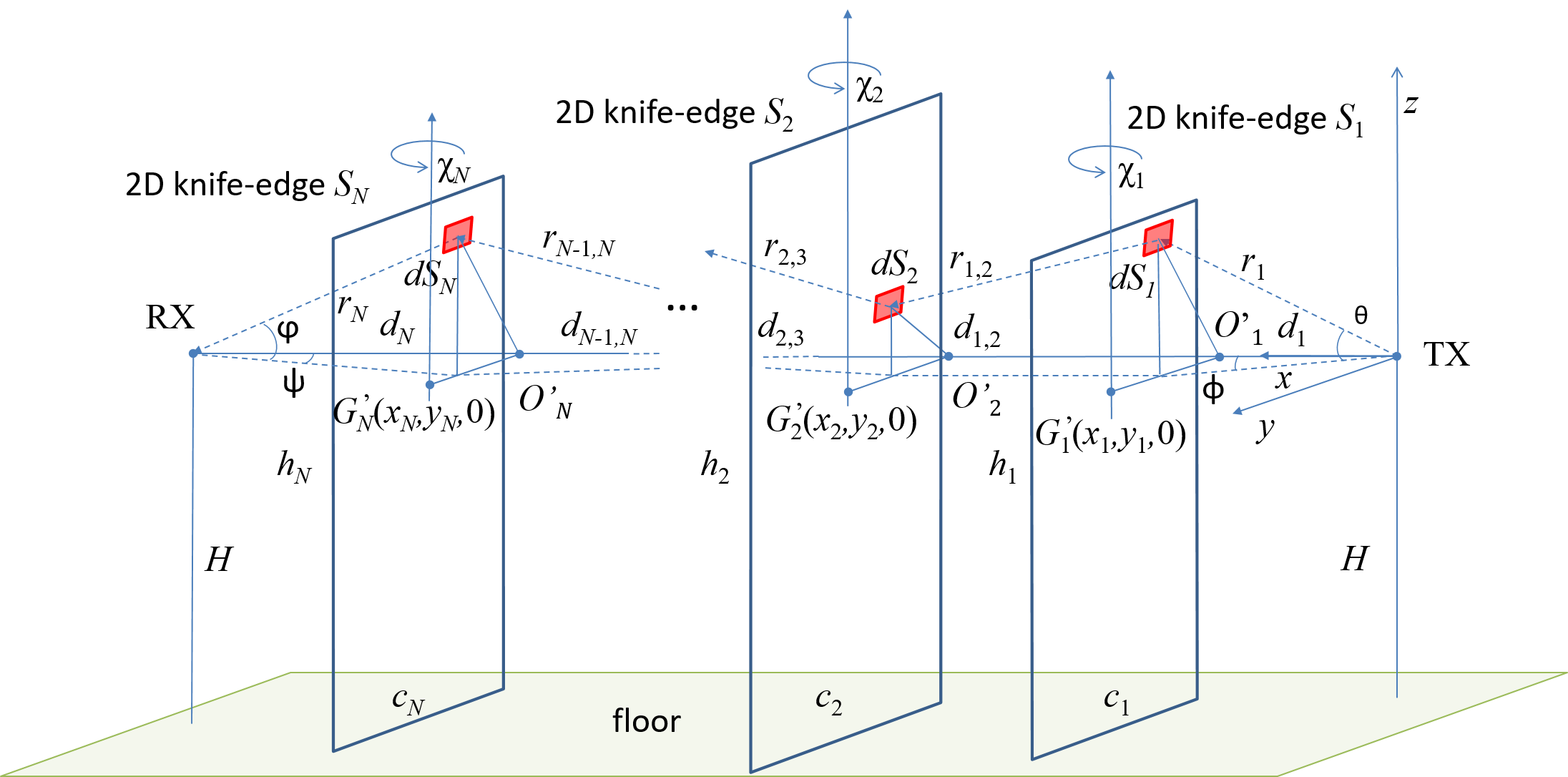}
\par\end{centering}
\caption{\label{fig:link}Single-link multi-target ($N>1$) scenario composed
by an horizontal single-link of length $d$, placed at distance $H$
from the floor, and $N$ different 2D equivalent knife-edge surfaces
$S_{1},S_{2},$...,$S_{N}$ corresponding to the targets $T_{1},T_{2},$...,$T_{N}$
located in $\mathbf{X}_{1},\mathbf{X}_{2},...,\mathbf{X}_{N}$, respectively.}

\vspace{-0.4cm}
\end{figure}

According to Fig. \ref{fig:link}, the multi-body scenario is a generalization
of the single-target case shown in Fig. \ref{fig:single}. $N$ knife-edge
surfaces stand vertically on the floor and are placed orthogonally
to the LOS path. They are numbered from $1$ up to $N$ according
to their increasing distances from the TX. Knife-edge positions are
identified by the column vectors $\mathbf{X}_{n}=[x_{n},y_{n}]^{T}$
that correspond to the projections $G_{n}^{'}=\left(x_{n},y_{n},0\right)$
of knife-edge barycenters $G_{n}=\left(x_{n},y_{n},z_{n}\right)$
with $z_{n}=h_{n}/2-H$. The positions of all targets are thus identified
by the column vector $\mathbf{X}=[\mathbf{X}_{1}^{T}\,\mathbf{X}_{2}^{T}\,...\,\mathbf{X}_{N}^{T}]^{T}$.
The single \emph{n-}th target $T_{n}$, with $1\leq n\leq N$, is
described by the rotating knife-edge surface $S_{n}$ having height
$h_{n}$, traversal size $c_{n}=c_{n}\left(\chi_{n}\right)$ with
constraint $a_{n}\leq c_{n}\leq b_{n}$ and orientation $\chi_{n}\in[-\pi,\pi]$
\emph{w.r.t.} the LOS path, with obvious meaning of these terms already
mentioned in Sect. \ref{subsec:Single-body-model}. All geometrical
and motion parameters are organized in the following column vectors:
$\mathbf{a}=[a_{1},\,a_{2},\,...,\,a_{N}]^{T}$, $\mathbf{b}=[b_{1},\,b_{2},\,...,\,b_{N}]^{T}$
collecting the knife-edges traversal sizes, $\mathbf{h}=[h_{1},\,h_{2},\,...,\,h_{N}]^{T}$
the target heights and $\boldsymbol{\chi}=[\chi_{1},\,\chi_{2},\,...,\,\chi_{N}]^{T}$
the corresponding orientations. Finally, $\forall\Delta x_{n},\Delta y_{n}\in\left[-B\:+B\right]$,
the column vector $\mathbf{\Delta X}=[\mathbf{\Delta X}_{1}^{T}\,\mathbf{\Delta X}_{2}^{T}\,...\,\mathbf{\Delta X}_{N}^{T}]^{T}$
tracks the involuntary/voluntary movements of the bodies in the position
interval $\left[-B\:+B\right]$ (supposed symmetrically arranged)
around the nominal position vector $\mathbf{X}$.

For $N>1$ targets, the knife-edge diffraction model (\ref{eq:dE_tot})
still holds true for each \emph{n-}th surface $S_{n}$, although equation
(\ref{eq:E_E0_tot-single}) is no longer valid. The LOS path is now
divided into $N+1$ segments of length equal to $d_{1},d_{1,2},d_{2,3},\ldots,d_{N-1,N},d_{N}$
with $d=d_{1}+\sum_{n=1}^{N-1}d_{n,n+1}+d_{N}$. Generalizing the
model of Sect. \ref{subsec:Single-body-model} for $N$ targets, the
term $r_{n,n+1}$ represents the distance between two consecutive
elementary areas $dS_{n}$ and $dS_{n+1}$ while the terms $r_{1}$
and $r_{N}$ represent the distance between the transmitter and $dS_{1}$,
and the distance between $dS_{N}$ and the receiver, respectively.
The \emph{n}-th elementary area $dS_{n}=d\xi_{n}d\varsigma_{n}$ is
located on the \emph{n-}th plane $S_{n}$ that is identified by its
position $\mathbf{X}_{n}$. The coordinate axes $\xi_{n}$ and $\varsigma_{n}$
(for clarity not shown in Fig. \ref{fig:link}) have the origin in
$O_{n}^{'}$ and are directed as the $y$ and $z$ axes. As an additional
hypothesis with respect to the ones of Sect. \ref{subsec:Single-body-model},
only forward propagation from TX to RX is considered with no backward
scattered waves between any surfaces $S_{n}$ and the RX (\ie both
single- and multiple-scattering effects between knife-edges are ignored).

In far field conditions, by assuming only forward propagation, $\forall T_{n}$
with $n=1,\ldots,N-1$, the elementary electric field $dE_{n+1}$
due to the diffraction effects caused by the elementary Huygens' source
of area $dS_{n}$, is computed at $dS_{n+1}$ by considering the distance
$r_{n,n+1}$ between the elements $dS_{n}$ and $dS_{n+1}$ according
to

\begin{equation}
dE_{n+1}=j\,\frac{dE_{n}\,dS_{n}}{\lambda r_{n,n+1}}\exp\left(-j\frac{2\pi r_{n,n+1}}{\lambda}\right)\,.\label{eq:dE_iterative}
\end{equation}
The electric field $E_{1},$ impinging on the first target, is

\begin{equation}
E_{1}=E_{0}\,\left(\frac{d}{r_{1}}\right)\exp\left(-j\frac{2\pi\left(r_{1}-d\right)}{\lambda}\right)\,,\label{eq:E_first}
\end{equation}
while the electric field $dE$, that is measured at the RX node and
generated by the area $dS_{N}$ of the target $N$ closest to the
RX is given by

\begin{equation}
dE=j\,\frac{dE_{N}\,dS_{N}}{\lambda r_{N}}\exp\left(-j\frac{2\pi r_{N}}{\lambda}\right)\,.\label{eq:E_last}
\end{equation}
Combining Eqs. (\ref{eq:dE_iterative}-\ref{eq:E_last}), it is

\begin{equation}
\begin{aligned}dE={} & j^{N}\frac{d}{\lambda^{N}\,r_{N}\,r_{N-1,N}\,...\,r_{1,2}\,r_{1}}\,E_{0}\cdot\\
{} & \cdot\exp\left\{ -j\frac{2\pi}{\lambda}\left(r_{N}+r_{N-1,N}+...+r_{1,2}+r_{1}-d\right)\right\} \cdot\\
{} & \cdot dS_{1}\,dS_{2}\,...\,dS_{N}\,.
\end{aligned}
\label{eq:dE_tot_N}
\end{equation}
To obtain the electric field $E$, Eq. (\ref{eq:dE_tot_N}) must be
integrated over the domain $S^{\left(c\right)}=\bigcup_{n=1}^{N}S_{n}^{\left(c\right)}$
where each region $S_{n}^{\left(c\right)}$ corresponds to the 2D
plane $\mathcal{P}_{n}\supset S_{n}$ that does not contain the points
of the knife-edge surface $S_{n}$. Eq. (\ref{eq:dE_tot_N}) becomes
now

\begin{equation}
\begin{aligned}\frac{E}{E_{0}}={} & j^{N}\int\limits _{S_{1}^{^{\left(c\right)}}}\int\limits _{S_{2}^{^{\left(c\right)}}}\cdots\int\limits _{S_{N}^{^{\left(c\right)}}}\frac{d}{\lambda^{N}\,r_{N}\,r_{N-1,N}\,...\,r_{1,2}\,r_{1}}\cdot\\
{} & \cdot\exp\left\{ -j\frac{2\pi}{\lambda}\left(r_{N}+r_{N-1,N}+...+r_{1,2}+r_{1}-d\right)\right\} \cdot\\
{} & \cdot dS_{1}\,dS_{2}\,...\,dS_{N}\,.
\end{aligned}
\label{eq:E_E0_tot_N}
\end{equation}

We now define $E^{\left(n\right)}$ as the value of the electric field
at the RX node when only one target, \ie the \emph{n-}th body or
obstacle, is present in the LOS area. Similarly, $E^{\left(n,m\right)}$
refers to the electric field at the receiver when only the \emph{n-}th
and \emph{m-}th obstacles out of $N$, are in the link area, and so
on. The notation $E^{\left(1,2,...,N\right)}$ thus highlights the
contributions of the $N$ targets to the link loss: this is the total
electric field $E$ at the receiver given by (\ref{eq:E_E0_tot_N}).
Considering the above definitions, Eq. (\ref{eq:E_E0_tot_N}) may
be rewritten to highlight the mutual interactions of the targets,
grouped by $\tbinom{N}{N-1}$ singles, $\tbinom{N}{N-2}$ pairs, $\tbinom{N}{N-3}$
triples and so on, as

\begin{equation}
\begin{aligned}\left(-1\right)^{N}\,\frac{E^{\left(1,2,..,N\right)}}{E_{0}}{} & =-1+\underset{\mathrm{singles}}{\underbrace{\sum_{n=1}^{N}\frac{E^{\left(n\right)}}{E_{0}}}}-\underset{\mathrm{pairs}}{\underbrace{\sum_{n=1}^{N-1}\sum_{m=n+1}^{N}\frac{E^{\left(n,m\right)}}{E_{0}}}}+\\
+{} & \underset{\mathrm{triples}}{\underbrace{\sum_{n=1}^{N-2}\sum_{m=n+1}^{N-1}\sum_{k=m+1}^{N}\frac{E^{\left(n,m,k\right)}}{E_{0}}}}+...+\\
{} & +\Psi(S_{1},...,S_{N})
\end{aligned}
\label{eq:E_E0_tot_N_iter}
\end{equation}
where the last term $\Psi(S_{1},...,S_{N})$: 
\begin{equation}
\begin{aligned}\Psi(S_{1},...,S_{N}){}{} & =j^{N}\int\limits _{S_{1}}\int\limits _{S_{2}}...\int\limits _{S_{N}}\frac{d}{\lambda^{N}\,r_{N}\,r_{N-1,N}\,...\,r_{1,2}\,r_{1}}\cdot\\
{} & \cdot\exp\left\{ -j\frac{2\pi}{\lambda}\left(r_{N}+r_{N-1,N}+...\right.\right.\\
{} & \left.\left....+r_{1,2}+r_{1}-d\right)\right\} \cdot dS_{1}dS_{2}...dS_{N}\,,
\end{aligned}
\label{eq:integ}
\end{equation}
is the integral computed over the composite domain defined by the
union $S^{\left(1,2,...,N\right)}=\bigcup_{n=1}^{N}S_{n}$ of the
$N$ rectangular knife-edge surfaces $S_{n}$. The knife edge surfaces
(Fig. \ref{fig:link}) have the following definitions: for $n=1$
it is $S_{1}=\left\{ \left(x,y,z\right)\in\mathbb{R^{\textrm{3}}}:\right.$
$x=x_{1}=d_{1}$, $y_{1}-c_{1}/2\leq$ $y$ $\leq y_{1}+c_{1}/2$,
$\left.-H\leq z\leq h_{1}-H\right\} $ while $\forall n=2,...,N$
it is also $S_{n}=\left\{ \left(x,y,z\right)\in\mathbb{R^{\textrm{3}}}:\right.$
$x=x_{n}=d_{1}+\sum_{i=1}^{n-1}d_{i,i+1}$, $y_{n}-c_{n}/2\leq$ $y$
$\leq y_{n}+c_{n}/2$, $\left.-H\leq z\leq h_{n}-H\right\} $.

Using (\ref{eq:E_E0_tot_N_iter}), for a generic number of targets
$N$, the electric field ratio $\frac{E^{\left(1,2,..,N\right)}}{E_{0}}$
due to $N$ obstructing bodies is composed by the single target contributions
$\frac{E^{\left(n\right)}}{E_{0}}$, for $n=1,...,N$ terms, the target
pairs, $\frac{E^{\left(n,m\right)}}{E_{0}}$, for $n=1,...,N-1,$$m=n+1,...,N,$
the triples, $\frac{E^{\left(n,m,k\right)}}{E_{0}}$, for $n=1,...,N-2$,
$m=n+1,...,N-1$, $k=m+1,...,N$, and so on, up to the contributions
of the $N-1$ target groups. Likewise PSBM and SBM, MBM gives valid
predictions when the targets are placed in the area $\mathcal{Y}$
near the LOS path, as defined in Sect. \ref{subsec:Single-body-model}.

When two bodies are in the area $\mathcal{Y}$, the received electric
field $E^{\left(1,2\right)}$ embeds the mutual effects of the two
targets $T_{1}$ (\ie $S_{1}$) and $T_{2}$ (\ie $S_{2}$) on the
radio propagation. $E^{\left(1,2\right)}$ is computed from the single-target
terms $E^{\left(1\right)}$ and $E^{\left(2\right)}$ as

\begin{equation}
\begin{aligned}\frac{E^{\left(1,2\right)}}{E_{0}}={} & -1+\frac{E^{\left(1\right)}}{E_{0}}+\frac{E^{\left(2\right)}}{E_{0}}+\Psi(S_{1},S_{2}).\end{aligned}
\label{eq:E_E0_tot_2}
\end{equation}
where the \emph{mixed} term, that depends on both knife-edges $S_{1}$
and $S_{2}$, is defined according to Eq. (\ref{eq:integ}) as
\begin{equation}
\begin{aligned}\Psi(S_{1},S_{2}){}{} & =-\int\limits _{S_{1}}\int\limits _{S_{2}}\frac{d}{\lambda^{2}\,r_{2}\,r_{1,2}\,r_{1}}\cdot\\
{} & \cdot\exp\left\{ -j\frac{2\pi}{\lambda}\left(r_{2}+r_{1,2}+r_{1}-d\right)\right\} dS_{1}dS_{2}\,.
\end{aligned}
\label{eq:integ_n2}
\end{equation}
In particular, from Eq. (\ref{eq:E_E0_tot_2}), the term $E^{\left(1\right)}$
quantifies the effect of the target $T_{1}$ alone in the link area
according to Eq. (\ref{eq:E_E0_tot-single}). It depends on the corresponding
target size $c_{1}$, the target height $h_{1}$, the link height
$H$ from the floor, and the distances $d_{1}$ and $d-d_{1}=d_{2}+d_{12}$
of the body $T_{1}$ from the TX an RX, respectively. Likewise, $E^{\left(2\right)}$
refers to the contributions of target $T_{2}$ only, according to
its target size $c_{2}$ and height $h_{2}$, the link height $H$
from the floor, and the distances $d_{1}+d_{12}$ and $d_{2}$ of
the body $T_{2}$ from the TX an RX, respectively.

For $N=1$, the proposed multi-body model (MBM) reduces to the single
body model (SBM) as expected since all singles, pairs, triples and
other high order terms of Eq. (\ref{eq:E_E0_tot_N_iter}) vanish except
for the term $\Psi(S_{1})$. The MBM model for $N=2$ targets (\ref{eq:E_E0_tot_2})
has been initially introduced in \cite{Rampa-2019} along with some
preliminary results. This dual-target model can be directly obtained
from the Eq. (\ref{eq:E_E0_tot_N}) or, equivalently, Eq. (\ref{eq:E_E0_tot_N_iter}).
Model comparisons are presented in Sect. \ref{subsec:Model-calibration}.

\subsection{Paraxial multi-body model (PMBM)}

\label{subsec:paraxial_sec}

For HPS applications, paraxial hypotheses are realistic only for small
target(s), namely for small enough $c_{i}$ and $h_{i}$ \emph{w.r.t.
}the path length $d$. The approximation also requires the subject
to move nearby the LOS path, with small enough $y_{i}$ and $z_{i}$,
or located in the central part of the LOS path. Carrier wavelength
$\lambda$ is also much smaller than the distances $x_{i}$ and $d-x_{i}$
(Sect. \ref{subsec:Paraxial-single-body}). Since the paraxial approximation
is useful in several applications, mostly outdoor, in the following
section, we will approximate the full Eqs. (\ref{eq:E_E0_tot_N})
or (\ref{eq:E_E0_tot_N_iter}) using paraxial assumptions. Such a
model will be labeled as PMBM (Paraxial MBM).

Based on the paraxial approximation, Eq. (\ref{eq:E_E0_tot_N}) becomes

\begin{equation}
\begin{aligned}\frac{E}{E_{0}}={} & \left(\frac{j}{2}\right)^{N}\int\limits _{S_{1}^{^{\left(c\right)}}}\int\limits _{S_{2}^{^{\left(c\right)}}}\cdots\int\limits _{S_{N}^{^{\left(c\right)}}}\\
{} & \frac{d\,d_{1,2}d_{2,3}...d_{N-1,N}}{\left(d_{1}+d_{1,2}\right)\left(d_{1,2}+d_{2,3}\right)...\left(d_{N-1,N}+d_{N}\right)}\cdot\\
\cdot{} & \exp\left\{ -j\frac{\pi}{2}\left(u_{1}^{2}+u_{2}^{2}+...+u_{N}^{2}-2\alpha_{1,2}u_{1}u_{2}...+\right.\right.\\
-{} & \left.\left.2\alpha_{N-1,N}u_{N-1}u_{N}\right)\right\} du_{1}du_{2}...du_{N}\cdot\\
\cdot{} & \exp\left\{ -j\frac{\pi}{2}\left(v_{1}^{2}+v_{2}^{2}+...+v_{N}^{2}-2\alpha_{1,2}v_{1}v_{2}...+\right.\right.\\
-{} & \left.\left.2\alpha_{N-1,N}v_{N-1}v_{N}\right)\right\} dv_{1}dv_{2}...dv_{N}.
\end{aligned}
\label{eq:E_E0_tot_N_app}
\end{equation}
In the Appendix \ref{sec:Appendix}, we show how to rewrite Eq. (\ref{eq:E_E0_tot_N_app})
to reveal the mutual interactions of targets as in Eq. (\ref{eq:E_E0_tot_N_iter}).

For the case of $N=2$ targets, Eq. (\ref{eq:E_E0_tot_N_app}) becomes
now analytically tractable. Using the formulation shown in Eq. (\ref{eq:E_E0_tot_2}),
adapted in Appendix \ref{sec:Appendix} for paraxial assumptions,
it is 
\begin{equation}
\begin{aligned}\frac{E^{\left(1,2\right)}}{E_{0}}={} & -1+\frac{E^{\left(1\right)}}{E_{0}}+\frac{E^{\left(2\right)}}{E_{0}}-\frac{1}{4}\frac{d\,d_{1,2}}{\left(d_{1}+d_{1,2}\right)\left(d_{1,2}+d_{2}\right)}\cdot\\
\cdot{} & \int_{-\unit{\sqrt{2}\mathit{H}}/R_{1}}^{+\sqrt{2}\left(h_{1}-H\right)/R_{1}}\int_{-\unit{\sqrt{2}\mathit{H}/\mathit{R_{2}}}}^{+\sqrt{2}\left(h_{2}-H\right)/R_{2}}\\
{} & \exp\left\{ -j\frac{\pi}{2}\left(u_{1}^{2}+u_{2}^{2}-2\alpha_{1,2}u_{1}u_{2}\right)\right\} du_{1}du_{2}\cdot\\
\cdot{} & \int_{\left(\sqrt{2}y_{1}-c_{1}/\sqrt{2}\right)/R_{1}}^{\left(\sqrt{2}y_{1}+c_{1}/\sqrt{2}\right)/R_{1}}\int_{\left(\sqrt{2}y_{2}-c_{2}/\sqrt{2}\right)/R_{2}}^{\left(\sqrt{2}y_{2}+c_{2}/\sqrt{2}\right)/R_{2}}\\
{} & \exp\left\{ -j\frac{\pi}{2}\left(v_{1}^{2}+v_{2}^{2}-2\alpha_{1,2}v_{1}v_{2}\right)\right\} dv_{1}dv_{2}\,,
\end{aligned}
\label{eq:E_Er_2_1_N}
\end{equation}
where $u_{1},u_{2}$,$v_{1},v_{2}$ and the constant terms $R_{1},R_{2}$
and $\alpha_{1,2}$ defined in the same Appendix \ref{sec:Appendix}.
Notice that, some approximated models are already available in the
literature \cite{lee1978path,vogler-1982} for the evaluation of the
extra attenuation due to multiple semi-infinite knife-edge surfaces.
These models can be obtained by using the paraxial approximation over
the semi-infinite domains representing the targets (\ie unlike the
\emph{finite} target size assumption adopted here); they are typically
effective in outdoor scenarios for the prediction of the propagation
loss over non-regular terrain profiles. The interested reader can
take a look at \cite{tzaras-2000} (and references therein) for a
brief discussion and comparisons.

\subsection{Additive models}

\label{subsec:EM-vs.-additive}

Based on the analysis of the previous sections, the term $\left|E^{\left(1,2,...,N\right)}/E_{0}\right|$
for $N$ targets can be used to evaluate the extra attenuation $A_{\textrm{dB}}^{\left(1,2,...,N\right)}$
with respect to the free-space (\ie unobstructed or empty) scenario
as
\begin{equation}
A_{\textrm{dB}}^{\left(1,2,...,N\right)}=-10\,\log_{10}\left|E^{\left(1,2,...,N\right)}/E_{0}\right|^{2}.\label{eq:dbatt}
\end{equation}
From Eq. (\ref{eq:E_E0_tot_N_iter}), it is apparent that the extra
attenuation terms $\left|\frac{E^{\left(1\right)}}{E_{0}}\right|,\left|\frac{E^{\left(2\right)}}{E_{0}}\right|,\,...,\left|\frac{E^{\left(N\right)}}{E_{0}}\right|$
alone or, equivalently, $A_{\textrm{dB}}^{\left(1\right)},\,A_{\textrm{dB}}^{\left(2\right)},...,\,A_{\textrm{dB}}^{\left(N\right)}$,
are not sufficient to evaluate $\left|\frac{E^{\left(1,2,..,N\right)}}{E_{0}}\right|$
since: \emph{i}) the phase relations between the terms $\frac{E^{\left(n\right)}}{E_{0}}$
are unknown, \emph{ii}) the terms $\frac{E^{\left(n,m\right)}}{E_{0}},\,\frac{E^{\left(n,m,k\right)}}{E_{0}},\,...,\,$
are not available, and \emph{iii}) the interaction terms between the
targets, that are expressed by the integral of the right side of Eq.
(\ref{eq:E_E0_tot_N_iter}), are not known as well. These facts prevent
the use of single-target measurements for the multiple target case.
According to these considerations, the additive hypothesis, namely
$A_{\textrm{dB}}^{\left(1,2,...,N\right)}=A_{\textrm{dB}}^{\left(1\right)}+A_{\textrm{dB}}^{\left(2\right)}+...+A_{\textrm{dB}}^{\left(N\right)}$,
that is generally exploited in various forms in the literature \cite{patwari10,wilson10},
is a rather superficial approximation. For the case of two targets
$(N=2)$, in Sect. \ref{subsec:Model-validation-with} an additive
SBM model is proposed where the individual extra attenuations $A_{\textrm{dB}}^{(1)}$,
$A_{\textrm{dB}}^{(2)}$,..., $A_{\textrm{dB}}^{(N)}$ follow the
SBM model described in Eq. (\ref{eq:E_E0_tot-single}). Limitations
of such representation are highlighted by a comparison with the MBM
and PMBM models.

\section{Physical-statistical multi-body model}

\label{sec:Physical-statistical-modeling}

In this section, we propose a true multi-target physical-statistical
model that relates the RSS to the link geometry ($d$, $H$), the
bodies locations $\mathbf{X}$, and their geometrical sizes (\ie
$\mathbf{a}$, $\mathbf{b}$ and $\mathbf{h}$). In addition to the
diffraction, or physical, component analyzed in Sect. \ref{sec:Diffraction-model},
the additional statistical component quantifies the uncertainty of
body movements, modelled here by small random voluntary/involuntary
motions $\Delta\mathbf{X}$ and rotations $\boldsymbol{\chi}$ around
the nominal position $\mathbf{X}$, as well as multipath fading, multiple
scattering between bodies, backward propagation effects, and other
random fluctuations, not included in the diffraction terms. For the
sake of simplicity, in the following sections, all geometrical parameters
defined in Sect. \ref{subsec:Multi-body-model} will be represented
by the compact set $\boldsymbol{\Lambda}=\left\{ \mathbf{a},\mathbf{b},\mathbf{h},d,H\right\} $.

Let $P$ be the RSS measurement performed by the receiver RX and expressed
in logarithmic scale (\ie usually in dBm), the power measurement
$P$ can be modeled as the sum of \emph{i}) the deterministic term
$P_{L}=10\,\log_{10}\left|E_{0}\right|^{2}$ due to the free-space
path-loss; \emph{ii}) the extra attenuation term $A_{\textrm{dB}}=A_{\textrm{dB}}^{\left(1,2,...,N\right)}$
in (\ref{eq:dbatt}) with respect to the free-space path-loss, caused
by the body-induced diffraction terms, and \emph{iii}) the Gaussian
random term $w$ that includes the lognormal multipath effects \cite{shadowing},
measurement noise and other random disturbances assumed normally distributed.
According to these assumptions, it is

\begin{equation}
P=\left\{ \begin{array}{ll}
P_{L}-A_{\textrm{dB}}^{\left(1,2,...,N\right)}+w & \;\textrm{iff }\exists\,\mathbf{X}_{n}\in\mathcal{Y}\\
P_{L}+w_{0} & \;\textrm{elsewhere}.
\end{array}\right.\label{eq:model-2}
\end{equation}
The free-space term $P_{L}$ is a constant that depends only on the
geometry of the scenario, the transmitted power, the gain and configuration
of the antennas, and the propagation coefficients \cite{knife_edge}.
The term $A_{\textrm{dB}}^{(1,2,...,N)}=A_{\textrm{dB}}\left(\mathbf{X},\Delta\mathbf{X},\boldsymbol{\chi},\boldsymbol{\Lambda}\right)$
is the extra attenuation, expressed in dB, due to the body-induced
diffraction with respect to the free-space scenario. Its argument
$\left|E^{\left(1,2,...,N\right)}/E_{0}\right|^{2}$ is computed using
(\ref{eq:E_E0_tot_N}) or (\ref{eq:E_E0_tot_N_iter}) for MBM and
(\ref{eq:E_E0_tot_N_app}) or (\ref{eq:E_E0_N_tot_app_iter}) for
PMBM. Propagation effects not included in the diffraction models (\ref{eq:E_E0_tot_N})
or (\ref{eq:E_E0_tot_N_iter}) are modeled by the Gaussian noise $w\sim\mathcal{N}$$\left(\Delta\mu_{\text{C}},\sigma_{0}^{2}+\Delta\sigma_{C}^{2}\right)$
with $\Delta\mu_{\text{C}}$ and $\Delta\sigma_{C}^{2}$ being the
residual stochastic body-induced multipath fading mean and variance
terms \cite{dfl,rampa2015letter}. $\sigma_{0}^{2}$ models the power
fluctuations induced by environmental changes outside the link area,
and not attributable to body movements around the LOS link.

For the empty scenario, where nobody is present in the link area,
namely the \emph{background} configuration, the RSS is simply modelled
as $P=P_{L}+w_{0}$ with $w_{0}\sim\mathcal{N}$$\left(0,\sigma_{0}^{2}\right)$.
Notice that in HPS systems, $\mu_{0}=\mathrm{E}_{w_{0}}\left[P\right]=P_{L}$
and $\sigma_{0}^{2}=\mathrm{Var_{w_{0}}}\left[P\right]$ can be evaluated
from field measurements during a calibration phase, when there are
no targets the link area. On the contrary, the presence of people
modifies both the mean $\mu_{1}\left(\mathbf{X}\right)=\textrm{\ensuremath{\mathrm{E_{\boldsymbol{\chi},\Delta\mathbf{X},\mathit{w}}}}}\left[P\right]$
and the variance $\sigma_{1}^{2}\left(\mathbf{X}\right)=\textrm{\ensuremath{\mathrm{\textrm{Var\ensuremath{\mathrm{_{\boldsymbol{\chi},\Delta\mathbf{X},\mathit{w}}}}}}}}\left[P\right]$
terms. Based on Eq. (\ref{eq:model-2}), the mean $\mu\left(P\right)$
and variance $\sigma^{2}\left(P\right)$ are defined as

\begin{equation}
\mu\left(P\right)=\left\{ \begin{array}{ll}
\mu_{1}\left(\left.\mathbf{X}\right|\boldsymbol{\Lambda}\right)=P_{L}+\Delta\mu\left(\left.\mathbf{X}\right|\boldsymbol{\Lambda}\right) & \;\textrm{iff }\exists\,\mathbf{X}_{n}\in\mathcal{Y}\\
\mu_{0}=P_{L} & \;\textrm{elsewhere}
\end{array}\right.\,\label{eq:mu}
\end{equation}
and

\begin{equation}
\sigma^{2}\left(P\right)=\left\{ \begin{array}{ll}
\sigma_{1}^{2}\left(\left.\mathbf{X}\right|\boldsymbol{\Lambda}\right)=\sigma_{0}^{2}+\Delta\sigma^{2}\left(\left.\mathbf{X}\right|\boldsymbol{\Lambda}\right) & \;\textrm{iff }\exists\,\mathbf{X}_{n}\in\mathcal{Y}\\
\sigma_{0}^{2} & \;\textrm{elsewhere}
\end{array}\right.\,\label{eq:sigma}
\end{equation}
where it is emphasized the dependency of $P$ from the position $\mathbf{X}_{n}$
of at least one target $T_{n}$ in the area $\mathcal{Y}$ and the
geometrical coefficients $\boldsymbol{\Lambda}$. The RSS average
$\Delta\mu\left(\mathbf{X}\right)=\mu_{1}\left(\mathbf{X}\right)-\mu_{0}$
and variance $\Delta\sigma^{2}\left(\mathbf{X}\right)=\sigma_{1}^{2}\left(\mathbf{X}\right)-\sigma_{0}^{2}$
increments are defined as
\begin{equation}
\Delta\mu\left(\left.\mathbf{X}\right|\boldsymbol{\Lambda}\right)=\Delta\mu_{C}-\mathrm{E_{\boldsymbol{\chi},\Delta\mathbf{X}}}\left[A_{\textrm{dB}}\left(\left.\mathbf{X}\right|\Delta\mathbf{X},\boldsymbol{\chi},\boldsymbol{\Lambda}\right)\right]\label{eq: delta P1 mean}
\end{equation}
and

\begin{equation}
\Delta\sigma^{2}\left(\left.\mathbf{X}\right|\boldsymbol{\Lambda}\right)=\Delta\sigma_{C}^{2}+\textrm{\ensuremath{\mathrm{\textrm{Var}{}_{\boldsymbol{\chi},\Delta\mathbf{X}}}}}\left[A_{\textrm{dB}}\left(\left.\mathbf{X}\right|\Delta\mathbf{X},\boldsymbol{\chi},\boldsymbol{\Lambda}\right)\right]\,.\label{eq:delta P1 sigma}
\end{equation}
The term $\textrm{\ensuremath{A_{\textrm{dB}}\left(\left.\mathbf{X}\right|\Delta\mathbf{X},\boldsymbol{\chi},\boldsymbol{\Lambda}\right)}}$
highlights the fact that, given the geometrical parameters $\boldsymbol{\Lambda}$
and the motion terms $\Delta\mathbf{X},\,\boldsymbol{\chi}$, the
extra attenuation is only a function of the positions $\mathbf{X}$
of the bodies. In the followings, we assume that the bodies are positioned
in $\mathbf{X}$ but each of them can slightly change its location
and posture making small random movements $\Delta\mathbf{X}_{n}$
with $\Delta x_{n},\Delta y_{n}\sim\mathcal{U}\left(-B,+B\right)$
and rotations $\chi_{n}\sim\mathcal{U}\left(-\pi,+\pi\right)$ around
the vertical axis. $\mathcal{U}\left(\alpha,\beta\right)$ indicates
the uniform distribution within the interval $\left[\alpha\:\beta\right]$
while, for each \emph{n}, the set$\left[-B\:+B\right]\times\left[-B\:+B\right]$
defines the 2D area around the nominal coordinate position $\mathbf{X}_{n}$
where the \emph{n}-th target can freely move. To determine eqs. (\ref{eq: delta P1 mean})
and (\ref{eq:delta P1 sigma}), the mean $\mathrm{E_{\boldsymbol{\chi},\Delta\mathbf{X}}}\left[\cdot\right]$
and the variance $\textrm{\ensuremath{\mathrm{\textrm{Var}{}_{\boldsymbol{\chi},\Delta\mathbf{X}}}}}\left[\cdot\right]$
are computed over the aforementioned uniform distribution of $\Delta\mathbf{X}$
and $\boldsymbol{\chi}$.

The residual body-induced multipath terms $\Delta\mu_{C}$ and $\Delta\sigma_{C}^{2}$
in (\ref{eq: delta P1 mean}) and (\ref{eq:delta P1 sigma}), respectively,
can be directly evaluated from field measurements performed during
the calibration phase. However, these terms are marginally influenced
by the specific body locations, as also shown in \cite{shadowing,rampa2015letter},
and are thus not relevant for HPS applications. On the contrary, the
diffraction term $A_{\textrm{dB}}\left(\left.\mathbf{X}\right|\Delta\mathbf{X},\boldsymbol{\chi},\boldsymbol{\Lambda}\right)$
provides a simple but effective method to predict the power perturbation
$\Delta\mu\left(\left.\mathbf{X}\right|\boldsymbol{\Lambda}\right)$
and $\Delta\sigma^{2}\left(\left.\mathbf{X}\right|\boldsymbol{\Lambda}\right)$
as a function of the body position and size.

\section{Model optimization and validation}

\label{subsec:Model-calibration}

To confirm the validity of the proposed multi-body models, several
EM simulations with the Feko software environment and on-field experiments
have been carried out according to the same link scenario sketched
in Fig. \ref{fig:link}. Feko\footnote{The commercial EM simulator designed by Altair Engineering Inc.}
implements time and frequency domain full-wave solvers (\eg MoM,
FDTD, FEM and MLFMM). For details about the solvers, the interested
reader may have a look at \cite{elsherbeni-2014antennabook}.

First of all, simulations have been carried out to compare the results
of the diffraction-based MBM and PMBM models shown in Sect. \ref{subsec:Multi-body-model}
and Sect. \ref{subsec:paraxial_sec}, respectively, with the ones
obtained with Feko. Next, we have compared the aforementioned models
against the RSS measurements obtained from IEEE 802.15.4 devices \cite{154E},
commonly used in industrial applications \cite{savazzi2016sensors}.
Considering the application to multi-body localization featuring $N=2$
targets, the MBM and PMBM model parameters, namely the geometrical
sizes (\ie $\mathbf{a}$, $\mathbf{b}$ and $\mathbf{h}$) of the
knife-edge surfaces, are optimized using a small subset of the experimental
data so that they could effectively model the obstructions induced
by the true targets. The proposed models using optimized sizes of
the knife-edges are then validated over different configurations,
where both targets move along the LOS link.

It is worth noticing that Feko simulations are related to PEC (Perfect
Electromagnetic Conductor) configurations to describe knife-edge targets.
On the contrary, MBM and PMBM 2D models assume perfectly absorbing
surfaces. The MBM and PMBM models also ignore important EM parameters
such as polarization, permittivity, conductivity, shape, radius of
curvature, and surface roughness \cite{Davis-et-al} that the Feko
simulator is able to tackle.

\begin{figure}[tp]
\begin{centering}
\includegraphics[clip,scale=0.4]{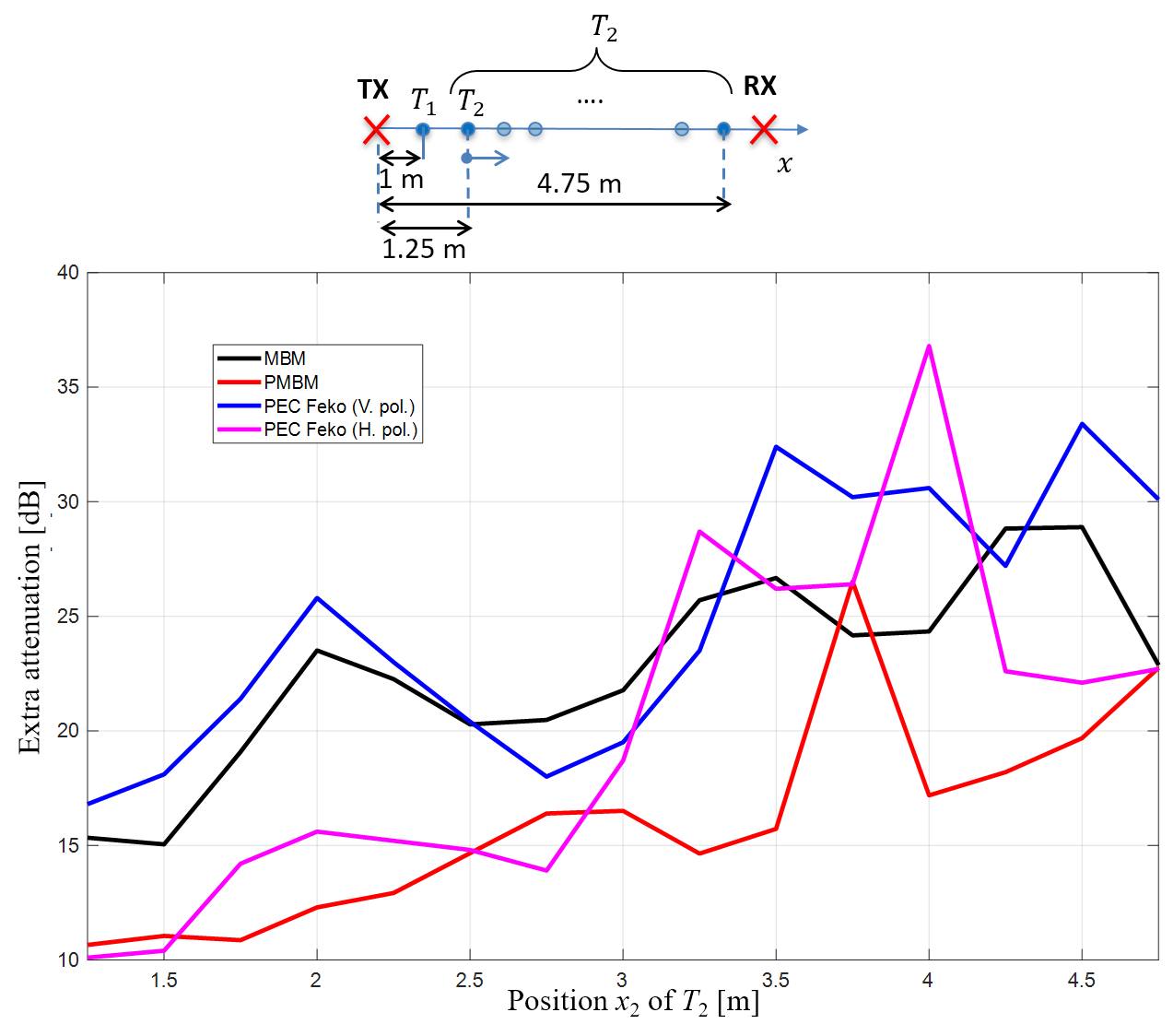}
\par\end{centering}
\caption{\label{fig:comp-along}Feko PEC simulations for vertically (blue line)
and horizontally (magenta line) polarized source vs. MBM (black line)
and PMBM (red line) $A_{\textrm{dB}}^{\left(1,2\right)}$ predictions
for targets $T_{1}$ and $T_{2}$ along the LOS path as shown in the
scenario on the top.}

\vspace{-0.4cm}
\end{figure}

\begin{figure}[tp]
\begin{centering}
\includegraphics[scale=0.4]{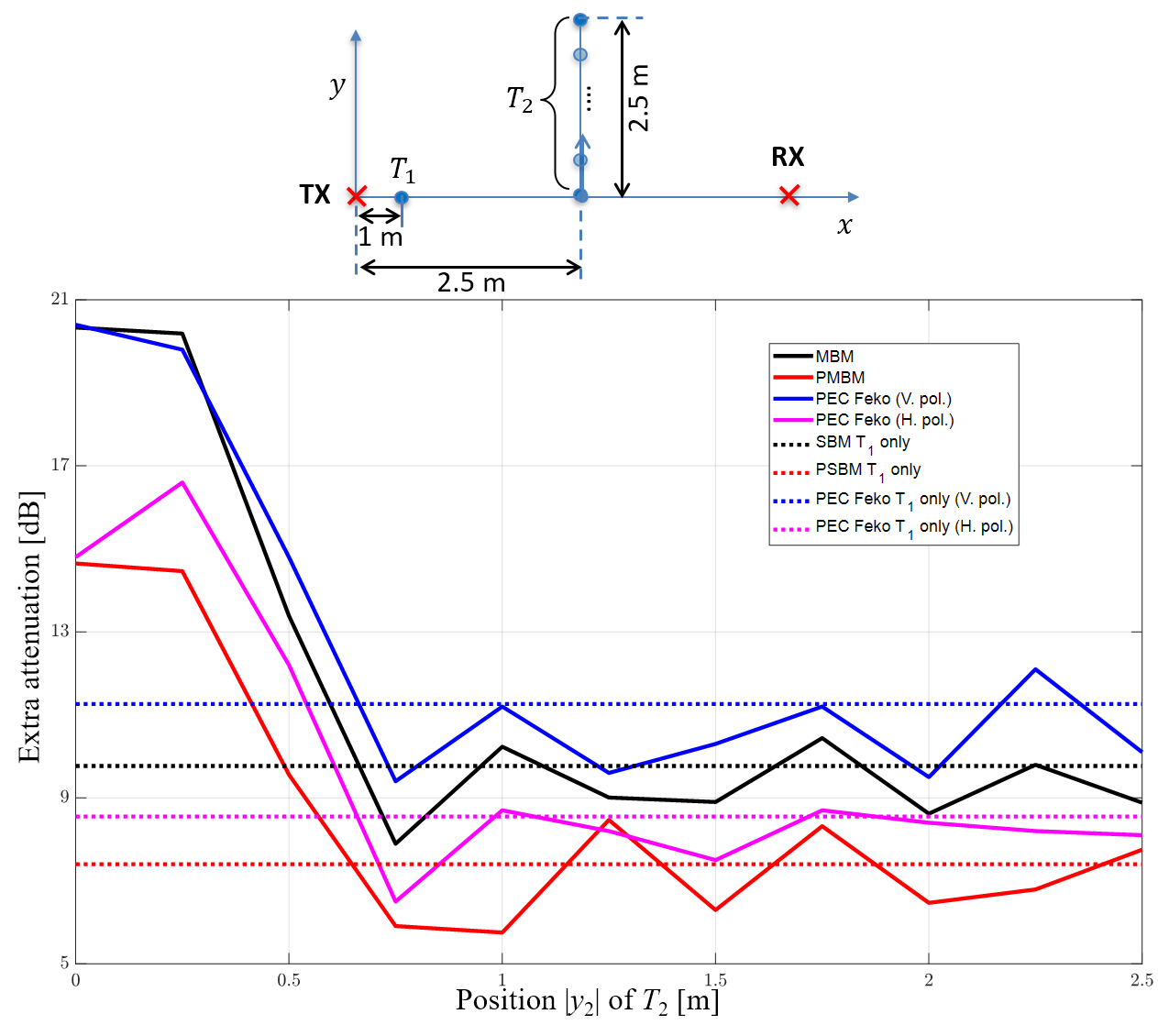}
\par\end{centering}
\caption{\label{fig:comp-across}Feko PEC simulations for vertically (blue
line) and horizontally (magenta line) polarized source vs. MBM (black
line) and PMBM (red line) $A_{\textrm{dB}}^{\left(1,2\right)}$ predictions
for the target $T_{1}$ along the LOS path and $T_{2}$ across the
LOS path as shown in the scenario on the top. The dotted lines (with
the same colors adopted for the dual-target cases) show the extra
attenuation predicted by the previous models/simulations due to the
presence of the target $T_{1}$ only.}

\vspace{-0.4cm}
\end{figure}

\subsection{Model comparison against EM simulations}

The MBM and PMBM models are evaluated and compared in this section
with the results from EM simulations. To simplify the EM simulation
complexity (mostly due to the long Feko runs), in what follows MBM
and PMBM models are compared in a dual-target scenario ($N=2$) only.
Fig. \ref{fig:comp-along} and \ref{fig:comp-across} show the predicted
values of the extra attenuation $A_{\textrm{dB}}^{\left(1,2\right)}=-10\,\log_{10}\left|E^{\left(1,2\right)}/E_{0}\right|^{2}$
computed according to the models described by eqs. (\ref{eq:E_E0_tot_2})
and (\ref{eq:E_Er_2_1_N}), namely MBM and PMBM, respectively. No
movements/rotations are allowed and both targets are placed in their
nominal positions: in both figures, the target $T_{1}$ is fixed in
the position $\mathbf{X}_{1}=[1,0]^{\mathrm{T}}$while the target
$T_{2}$ changes its positions along and across the LOS path. In Fig.
\ref{fig:comp-along}, $T_{2}$ is placed in $\mathbf{X}_{2}=[x_{2},0]^{\mathrm{T}}$
and moves along the LOS path with $1<x_{2}<d$ and $0.25$ m increments
while in Fig. \ref{fig:comp-across} the target $T_{2}$ is placed
in $\mathbf{X}_{2}=[d/2,y_{2}]^{\mathrm{T}}$ with $-2.5<y_{2}<2.5$
m and $0.25$ m increments thus crossing the LOS path in the middle.
For symmetry reasons, the results depends only on the distance $|y_{2}|$
from the LOS line.

In these figures, we compare the results of the MBM and PMBM against
the ones of the EM simulations obtained with Feko using vertically
and horizontally polarized sources (\ie Feko V. pol. and H. pol.,
respectively). For these simulations, a 2D PEC surface have been used
as target instead of the absorbing one defined in Sect. \ref{sec:Diffraction-model}
for MBM and PMBM, but with the same physical dimensions adopted for
Fig. \ref{fig:layout}. Target $T_{1}$ and $T_{2}$ have the same
size $c_{1}=c_{2}=0.55$ m and height $h_{1}=h_{2}=1.80$ m while
$H=0.90$ m and $d=5.0$ m. The average error $\epsilon_{\textrm{MBM}}$
(and $\epsilon_{\textrm{PMBM}}$) between MBM (and PMBM) and Feko
PEC V. pol. values in dB are summarized in Tab. \ref{tab:Error-values},
where the corresponding standard deviation values $\sigma_{\textrm{MBM}}$
(and $\sigma_{\textrm{PMBM}}$) are also included. The sign in the
average errors $\epsilon_{\textrm{MBM}}$ and $\epsilon_{\textrm{PMBM}}$
indicates that the results of the MBM and PMBM models underestimate
in average the Feko results.

\begin{table}
\caption{\label{tab:Error-values}Average errors and standard deviation values
in the LOS area obtained by MBM and PMBM predictions \emph{w.r.t.}
Feko PEC V. pol. simulations.}

\centering{}%
\begin{tabular}{|c|c|c|c|c|}
\hline 
LOS area & $\epsilon_{\textrm{MBM}}$ & $\sigma_{\textrm{MBM}}$ & $\epsilon_{\textrm{PMBM}}$ & $\sigma_{\textrm{PMBM}}$\tabularnewline
\hline 
\hline 
Along (Fig. \ref{fig:comp-along}) & $-2.1$ dB & $3.4$ dB & $-8.7$ dB & $4.4$ dB\tabularnewline
\hline 
Across (Fig. \ref{fig:comp-across}) & $-1.0$ dB & $0.7$ dB & $-4.0$ dB & $1.5$ dB\tabularnewline
\hline 
\end{tabular}
\end{table}

Along the LOS positions of Fig. \ref{fig:comp-along}, the average
errors are $\epsilon_{\textrm{MBM}}=-2.1$ dB and $\epsilon_{\textrm{PMBM}}=-8.7$
dB, respectively, while the corresponding standard deviations are
$\sigma_{\textrm{MBM}}=3.4$ dB and $\sigma_{\textrm{PMBM}}=4.4$
dB, as well. As already noted, while the MBM trend is similar to the
one due to the Feko predictions, the PMBM values introduce too many
large errors. These effects are also apparent in Fig. \ref{fig:comp-across}
where the target $T_{2}$ move along the orthogonal line that crosses
the LOS in the middle at $x_{2}=2.5$ m. Even in this case, the MBM
trend is similar to the PEC vertical polarized curve predicted by
the Feko software with an maximum error in the order of $1\div2$
dB. In particular, $\epsilon_{\textrm{MBM}}=-1.0$ dB, $\epsilon_{\textrm{PMBM}}=-4.0$
dB, $\sigma_{\textrm{MBM}}=0.7$ dB and $\sigma_{\textrm{PMBM}}=1.5$
dB considering all positions across the LOS line. In Fig. \ref{fig:comp-across},
the results corresponding to the models for the single target $T_{1}$
are superimposed to the ones predicted by the MBM/PMBM models and
the ones obtained with Feko. When target $T_{2}$ moves away from
the LOS path the effects due to this target vanish, while the extra
attenuation can be well predicted using only the single target body
model for $T_{1}$. The difference between the values predicted by
the MBM (and PMBM) model and the Feko simulations are due to the fact
that the former employs absorbing targets and neglects any polarization
while the latter exploits metallic plates and includes the polarization
effects. The positions with larger differences (\ie up to $5\div6$
dB) are visible in the right part of Fig. \ref{fig:comp-along} where
the target $T_{2}$ is very close to the receiver and the metallic
nature of the PEC is more evident. Even if some positions of Fig.
\ref{fig:comp-along} show some relevant differences between MBM and
Feko results, the general trend is maintained. On the contrary, differences
between the MBM (and PMBM) model and the Feko simulations are smaller
for off-axis placement of $T_{2}$ as shown in Fig. \ref{fig:comp-across}.
It is also worth noticing in Fig. \ref{fig:comp-along} the strong
differences between vertically and horizontally polarized results
obtained with Feko.

\begin{figure*}[!tp]
\begin{centering}
\includegraphics[scale=0.4]{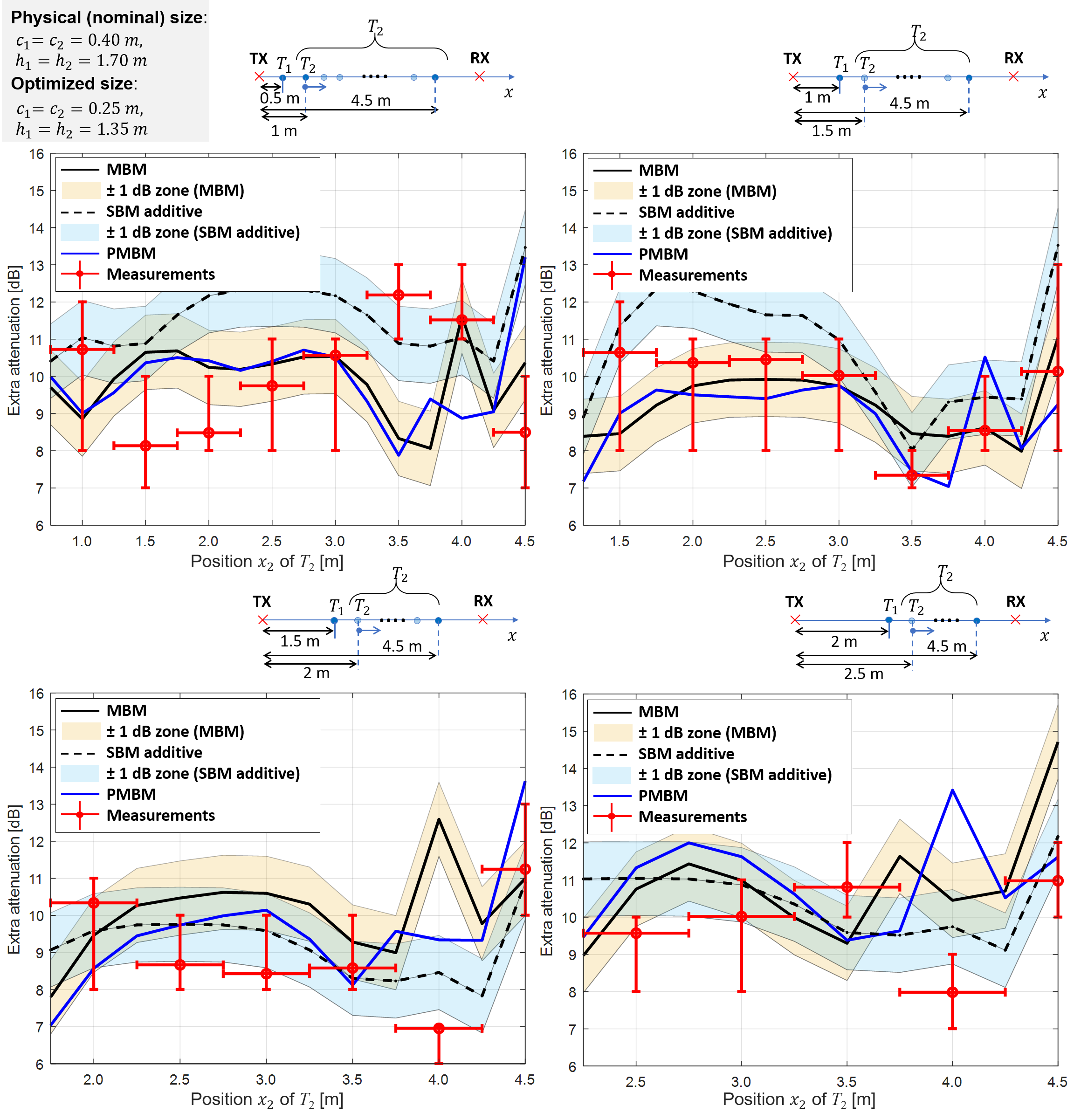}
\par\end{centering}
\caption{\label{fig:measurements}Predicted extra attenuation values using
MBM (black line), PMBM (blue line) and additive SBM (black dashed
line) against the measured ones (red cross, with vertical and horizontal
error bars) considering $4$ motion scenarios featuring $N=2$ targets.
The $\pm1$ dB uncertainty zones are shown around the additive SBM
(light blue area) and the MBM (beige area) results. Each scenario
is depicted in the corresponding sub-figure.}

\vspace{-0.4cm}
\end{figure*}

\subsection{Model optimization and experiments}

\label{subsec:Model-optimization}

Localization, and, in general, HPS systems, rely on several calibration
steps that include off-line measurements stages to collect ground-truth
measurements of radio propagation. Detection algorithms based on finger-printing
approaches \cite{savazzi2016magazine} require also the presence of
the target(s) in known locations, namely landmark points, to perform
RF measurements with target(s) inside the monitored area. These steps
are very time consuming and error-prone since each RF node needs to
perform noisy RSS signal measurements for the assessment of body-induced
alterations of the radio propagation with the target(s) placed in
the selected landmarks (\ie 2-4 landmarks/sqm depending on the network
placement). In addition, systematic errors can be also introduced
in the experimental setup during target positioning. Moreover, modifications
of the environment involve frequent recalibration steps to maintain
a good level of localization accuracy. In what follows, we investigate
the use of the multi-body models of Sect. \ref{sec:Physical-statistical-modeling}
optimized to replace, or simplify, such calibration phases.

Model optimization, or tuning, is implemented during an initial stage
to estimate the MBM and PMBM model hyper-parameters, the geometrical
sizes $\mathbf{a}$, $\mathbf{b}$ and $\mathbf{h}$. We thus collect
a small set of measurements in selected landmark points: these are
used to tune both MBM and PMBM physical parameters, namely the \emph{EM-equivalent}
height $h_{n}$ and width $c_{n}\in[a_{n},b_{n}]$ of the $n$-th
deployed subject. Using these optimized hyperparameters, the MBM and
the PMBM representations can be finally adopted to predict body effects
at arbitrary positions, replacing conventional calibration stages
\cite{savazzi2016magazine}.

RSS data have been collected in a large hall by using two IEEE 802.15.4-compliant
radio devices, based on the NXP JN5148 SoC (System-on-Chip) \cite{NXP},
that are employed as transceiver (TX/RX) nodes. Link geometry is $H=0.9$
m and $d=5$ m for all cases. Each device is equipped with 2 dBi vertical
monopole antennas and is programmed to send IEEE 802.15.4 standard
frames at frequency $f_{c}=2.486$ GHz (corresponding to IEEE 802.15.4
channel $26$ \cite{154E}). The receiver decodes each frame to extract
a measurement of RSS, namely the digital RSS Indicator (RSSI) with
8-bit resolution \cite{RSSI-calibration} and $1$ dB quantization.
Two set of measurements have been gathered: the first dataset, acquired
in the empty scenario, has been used to compute the reference values
$\mu{}_{0}$ and $\sigma_{0}^{2}$, while the second one has been
recorded with two targets are placed in the link area and move concurrently.
The second data set has been used to measure $\mu_{1}\left(\mathbf{X}\right)$
and $\sigma_{1}^{2}\left(\mathbf{X}\right)$ for some known landscape
positions $\mathbf{X}$. In particular, the subject $T_{1}$ was placed
at $\mathbf{X}_{1}=[2.5,0]^{\mathrm{T}}$ from the transmitter, the
second subject $T_{2}$ covered $4$ different positions, $\mathbf{X}_{2}=[x_{2},0]^{\mathrm{T}}$,
at distances $x_{2}=3$ m, $x_{2}=3.5$ m, $x_{2}=4$ m and $x_{2}=4.5$
m, respectively. These locations have been used to optimize the EM-equivalent
geometrical dimensions $c_{1},c_{2},h_{1},h_{2}$ of the individual
subjects using a Non-linear Least Squares (NLS) approach as in \cite{Rampa-2019-calib}.
The optimization is iterative and uses the nominal (or physical) size
of the targets as initial inputs, namely $c_{1}=c_{2}=0.4$ m and
$h_{1}=h_{2}=1.7$ m (neglecting head). Optimized EM-equivalent target
size are $c_{1}=c_{2}=0.25$ m and $h_{1}=h_{2}=1.35$ m. The EM-equivalent
size is about $20\%\div40\%$ smaller than the physical size of the
target as also shown in \cite{conducting_cylinder}: this might be
due to the fact that body-induced extra attenuations are mostly due
to torso and legs, while the effects caused by head and arms are negligible,
although they are responsible for some small RSSI fluctuations.

\subsection{Model comparison with measurements}

\label{subsec:Model-validation-with}

Based on the experimental set-up described in the previous section,
we compare in Fig. \ref{fig:measurements} the predicted extra attenuation
$\mathrm{E_{\boldsymbol{\chi},\Delta\mathbf{X}}}\left[A_{\textrm{dB}}^{(1,2)}\left(\left.\mathbf{X}\right|\Delta\mathbf{X},\boldsymbol{\chi},\boldsymbol{\Lambda}\right)\right]$
using the MBM (black line) and PMBM (blue line) models against the
field measurements (red cross). RSSI measurements, averaged over a
period of $1$ minute, are indicated in Fig. \ref{fig:measurements}
as red cross markers. Horizontal error bars are also depicted to account
for positioning inaccuracies ($\pm0.25$ m) during the tests while
vertical bars and dots indicate the max-min and mean RSSI values,
respectively. Solid lines show the predicted MBM and PMBM extra attenuations
using the optimized EM-equivalent parameters. For prediction, we consider
two knife-edges placed along the LOS path in fixed positions ($B=0$)
with no rotation. In particular, subject $T_{1}$ initially placed
at $\mathbf{X}_{1}=[0.5,0]^{\mathrm{T}}$ from the transmitter covers
a distance of $1.5$ m in the direction of the RX (thus stopping at
$\mathbf{X}_{1}=[2,0]^{\mathrm{T}}$ from the TX). Subject $T_{2}$
moves towards the receiver and stops at $0.5$ m from the device. 

To highlight the comparative analysis with measurements, we neglect
the residual stochastic body-induced multipath fading terms by assuming
$\Delta\sigma_{C}^{2}=0$ dB and $\Delta h_{C}=0$ dB, so that Eq.
(\ref{eq: delta P1 mean}) reduces to $-\Delta\mu\left(\mathbf{X}\right)=\mathrm{E_{\boldsymbol{\chi},\Delta\mathbf{X}}}\left[A_{\textrm{dB}}^{(1,2)}\left(\left.\mathbf{X}\right|\Delta\mathbf{X},\boldsymbol{\chi},\boldsymbol{\Lambda}\right)\right]=A_{\textrm{dB}}^{(1,2)}\left(\left.\mathbf{X}\right|\boldsymbol{\Lambda}\right)$.
Besides MBM and PMBM models, based on the discussion in Sect. \ref{subsec:EM-vs.-additive},
for the same settings, we have also compared the additive hypothesis,
namely $A_{\textrm{dB}}^{\left(1,2\right)}=A_{\textrm{dB}}^{\left(1\right)}+A_{\textrm{dB}}^{\left(2\right)}$.
In particular, the additive SBM approximation is depicted in dashed
lines: in this case, the predicted extra attenuation reduces to $-\Delta\mu\left(\mathbf{X}\right)=\mathrm{E_{\chi_{1},\Delta\mathbf{X}_{1}}}\left[A_{\textrm{dB}}^{(1)}\left(\left.\mathbf{X}\right|\Delta\mathbf{X},\boldsymbol{\chi},\boldsymbol{\Lambda}\right)\right]+\mathrm{E_{\chi_{2},\Delta\mathbf{X}_{2}}}\left[A_{\textrm{dB}}^{(2)}\left(\left.\mathbf{X}\right|\Delta\mathbf{X},\boldsymbol{\chi},\boldsymbol{\Lambda}\right)\right]=A_{\textrm{dB}}^{(1)}\left(\left.\mathbf{X}\right|\boldsymbol{\Lambda}\right)+A_{\textrm{dB}}^{(2)}\left(\left.\mathbf{X}\right|\boldsymbol{\Lambda}\right)$
where $A_{\textrm{dB}}^{(1)}\left(\left.\mathbf{X}\right|\boldsymbol{\Lambda}\right)$
and $A_{\textrm{dB}}^{(2)}\left(\left.\mathbf{X}\right|\boldsymbol{\Lambda}\right)$
follow the SBM model described in Sect. \ref{subsec:Single-body-model}
and use the same optimized EM-equivalent target size as that of the
MBM model.

In Fig. \ref{fig:measurements}, the shaded areas indicate the $\pm1$
dB uncertainty zone around the mean values predicted by the MBM (\ie
beige area) and the additive SBM (\ie light blue area) models. Assuming
Gaussian distributions, for each uncertainty zone, the 68.3\% of the
related events fall inside each shaded area.

The discrepancies between the proposed models and the measurements
may be large for some target positions, configurations and scenarios.
This is due to some approximations adopted for the MBM/PMBM and SBM/PSBM
models that are summarized here: \emph{i}) the diffraction models
assume a 2D knife-edge approximation to describe each 3D target. In
addition, each knife-edge surface is assumed to be completely absorbing
\cite{rampa2017em} while this is just an approximation for the human
body \cite{kibret2015characterizing}; \emph{ii}) the multi-body diffraction
model derived here assumes only forward propagation from the transmitter
to the receiver and ignores reverberations among targets; \emph{iii})
the diffraction model, being based on the scalar diffraction theory,
does not include polarization effects; \emph{iv}) the EM composition
of the targets is not considered. 

In addition, there are also some noisy effects during the RSS measurement:
\emph{v}) the human bodies are never fixed in specific positions but,
while standing, they have both voluntary and/or involuntary movements
around the nominal position due to movements of the legs, torso, arms
and head. This experimental fact can introduce variations in the order
of $2\div3$ dB for the single target case \cite{rampa2017em}\emph{;
vi}) due to the complex structure of the human body and the difficulties
to measure the true position of the body \cite{rampa2017em,Rampa-2019-calib},
the nominal (measured) position of the target is only approximately
known (with an error in the order of $10\div15$ cm); \emph{vii})
the measurements are taken in real scenarios where some multipath
effects are present. These effects are modeled as a lognormal distributed
noise even if this an approximated behavior \cite{shadowing}; \emph{viii})
the RSS values measured by the IEEE 802.15.4 radio devices are corrupted
by measurement noise and non-linearity effects \cite{RSSI-calibration}
in the order of $0.5\div1.5$ dB, depending on the adopted device.
Offset errors can be compensated due to the fact that the extra attenuation
is measured \emph{w.r.t.} the free-space scenarios but non-linearity
and quantization cannot. Moreover, the radios are subject to some
other artifacts since, at very low RSSI, namely when more than one
target is near the LOS path, some packets are lost due to body shadowing
effects \cite{rampa2015letter,dfl}; these target configurations introduce
some high attenuated RSSI samples, mainly near the RX or the TX.

In Fig. \ref{fig:measurements}, the MBM and PMBM predictions track
the average RSSI measurements with an average error of about $4$
dB and $6$ dB, respectively. Notice that most of the mismatches between
the models and the measurements are observed in correspondence to
one of the targets (or both) near the transmitter or the receiver.
In such cases, the RSSI measurements might be also affected by larger
communication errors due to the increase of body-induced extra attenuation.
Fluctuations of RSSI around the average value cause residual stochastic
terms that can be quantified as a variance term $\Delta\sigma_{C}^{2}\simeq3$
dB and negligible mean $\Delta h_{C}\simeq0$ dB: these are mostly
due to voluntary/involuntary movements of the bodies around their
nominal positions. The additive SBM model is generally less effective
compared with MBM, particularly when the mixed terms in Eq. (\ref{eq:integ_n2}),
caused by the interacting targets $T_{1}$ and $T_{2}$, could not
be neglected. This is for example the case when both subjects equally
contribute to the extra attenuation, \ie for subject $T_{1}$ placed
at $0.5$ m, $\mathbf{X}_{1}=[0.5,0]^{\mathrm{T}}$, and $1$ m, $\mathbf{X}_{1}=[1,0]^{\mathrm{T}}$,
from the transmitter.

\section{Conclusions}

\label{sec:Conclusions}

Based on the electromagnetic (EM) scalar diffraction theory, the paper
proposes for the first time an ad-hoc physical-statistical model to
describe the fluctuations of the radio signal caused by the presence
of an arbitrary number of targets between the transmitter and receiver.
The analytical model has been specifically tuned and optimized to
predict the effects of multiple bodies placed near the link area.
Therefore, it is instrumental to Device-Free Localization (DFL) applications
and Human Presence-aware Systems (HPS), including people access monitoring
and counting. The model results have been validated experimentally
by some field tests using real industrial wireless devices and EM
simulations as well, for comparative analysis. The proposed multi-body
model is able to predict the Received Signal Strength (RSS) measurements
accounting for the size, orientation, small movements, and positions
of the targets. In addition, it overcomes some restrictions of the
existing multi-body models, based on the linear superposition of the
subject effects, thus showing improved prediction accuracy.

\section*{Appendix}

\label{sec:Appendix}

In what follows, we resort to a formulation of Eq. (\ref{eq:E_E0_tot_N_app})
similar to that used in Eq. (\ref{eq:E_E0_tot_N_iter}). It is

\begin{equation}
\begin{aligned}\left(-1\right)^{N}\,\frac{E^{\left(1,2,..,N\right)}}{E_{0}}{} & =-1+\sum_{n=1}^{N}\frac{E^{\left(n\right)}}{E_{0}}-\sum_{n=1}^{N-1}\sum_{m=n+1}^{N}\frac{E^{\left(n,m\right)}}{E_{0}}+\\
+{} & \sum_{n=1}^{N-2}\sum_{m=n+1}^{N-1}\sum_{k=m+1}^{N}\frac{E^{\left(n,m,k\right)}}{E_{0}}+...+\\
{} & +\widetilde{\Psi}(S_{1},...,S_{N})
\end{aligned}
\label{eq:E_E0_N_tot_app_iter}
\end{equation}
with last term $\widetilde{\Psi}(S_{1},...,S_{N})$: 
\begin{equation}
\begin{aligned}\widetilde{\Psi}(S_{1},...,S_{N}){}{} & =\left(\frac{j}{2}\right)^{N}\int\limits _{S_{1}}\int\limits _{S_{2}}...\int\limits _{S_{N}}\\
{} & \frac{d\,d_{1,2}d_{2,3}...d_{N-1,N}}{\left(d_{1}+d_{1,2}\right)\left(d_{1,2}+d_{2,3}\right)...\left(d_{N-1,N}+d_{N}\right)}\cdot\\
{} & \exp\left\{ -j\frac{\pi}{2}\left(u_{1}^{2}+u_{2}^{2}+...+u_{N}^{2}-2\alpha_{1,2}u_{1}u_{2}...+\right.\right.\\
{} & \left.\left.2\alpha_{N-1,N}u_{N-1}u_{N}\right)\right\} du_{1}du_{2}...du_{N}\cdot\\
{} & \exp\left\{ -j\frac{\pi}{2}\left(v_{1}^{2}+v_{2}^{2}+...+v_{N}^{2}-2\alpha_{1,2}v_{1}v_{2}...+\right.\right.\\
{} & \left.\left.2\alpha_{N-1,N}v_{N-1}v_{N}\right)\right\} dv_{1}dv_{2}...dv_{N}.
\end{aligned}
\label{eq:integ-1}
\end{equation}
The variables $u_{1},...,u_{N},v_{1},...,v_{N}$ are obtained from
the corresponding local coordinates $\xi_{1},...,\xi_{N},\varsigma_{1},...,\varsigma_{N}$
by using, for each \emph{n}-th term, the following substitution rules:
$u_{n}=\xi_{n}\tfrac{\sqrt{2}}{R_{n}}$ and $v_{n}=\varsigma_{n}\tfrac{\sqrt{2}}{R_{n}}$.
The constant terms $R_{n}$ and $\alpha_{n,n+1}$ are related to the
the wavelength $\lambda$ and the geometric positions of the knife-edges
with respect to the LOS path. The terms $R_{n}$ are similar to the
the Fresnel's radius $R$ for the single-target case. These constants
are given by

\begin{equation}
\frac{1}{R_{n}^{2}}=\left\{ \begin{aligned}{} & \frac{1}{\lambda}\left(\frac{1}{d_{1}}+\frac{1}{d_{12}}\right) & \textrm{for }n & =1\\
{} & \frac{1}{\lambda}\left(\frac{1}{d_{n-1,n}}+\frac{1}{d_{n,n+1}}\right) & \textrm{for }n & =2,...,N-1\\
{} & \frac{1}{\lambda}\left(\frac{1}{d_{N}}+\frac{1}{d_{N-1,N}}\right) & \textrm{for }n & =N
\end{aligned}
\right.\label{eq:generalized_R}
\end{equation}
while coefficients $\alpha_{n,n+1}$ are defined, $\textrm{for }n=1,...,N-1$,
as 
\begin{equation}
\alpha_{n,n+1}=\frac{R_{n}\,R_{n+1}}{\lambda\,d_{n,n+1}}\,.\label{eq:alpha}
\end{equation}
For the single-target case, it is $R_{1}=R$ while all coefficients
$\alpha_{n,n+1}$ vanish. In addition, it is trivial to verify that
Eq. (\ref{eq:E_E0_tot_N_iter}) simply reduces to Eq. (\ref{eq:E_E0_tot-single})
while Eq. (\ref{eq:E_E0_N_tot_app_iter}) simplifies to (\ref{eq:E_E0_approx-single}).
For $N=2$, Eq. (\ref{eq:E_E0_tot_N_iter}) reduces to the dual-target
case (\ref{eq:E_E0_tot_2}) and proves the results \cite{nicoli2016eusipco}
applied for localization purposes.

\bibliographystyle{IEEEtran}
\bibliography{DFLBIB}

\end{document}